\documentclass[submission,copyright,creativecommons]{eptcs}
\providecommand{\event}{FMAS 2021} 
\usepackage{breakurl}             
\usepackage{underscore}           
\usepackage{graphicx}
\usepackage{array}
\usepackage{amssymb}
\usepackage{stmaryrd}
\usepackage{subcaption}
\usepackage{tikz}
\usepackage{float}
\usepackage{url}
\usepackage{doi}
\usepackage{circle}
\usepackage{upgreek}
\usepackage[all]{xy}

\usepackage[textsize=normalsize,textwidth=3cm]{todonotes}

\title{Towards Partial Monitoring: It is Always too Soon to Give Up\thanks{Cardoso's work supported by Royal Academy of Engineering under the Chairs in Emerging Technologies scheme.}}
\author{Angelo Ferrando
\institute{University of Genova\\ Genova, Italy}
\email{angelo.ferrando@unige.it}
\and
Rafael C. Cardoso
\institute{The University of Manchester\\
Manchester, United Kingdom}
\email{rafael.cardoso@manchester.ac.uk}
}
\def\titlerunning{Towards Partial Monitoring: It is Always too Soon to Give Up}
\def\authorrunning{A. Ferrando and R. C. Cardoso}
\begin{document}
\maketitle

\begin{abstract}

Runtime Verification is a lightweight formal verification technique. It is used to verify at runtime whether the system under analysis behaves as expected. The expected behaviour is usually formally specified by means of properties, which are used to automatically synthesise monitors. A monitor is a device that, given a sequence of events representing a system execution, returns a verdict symbolising the satisfaction or violation of the formal property. Properties that can (resp. cannot) be verified at runtime by a monitor are called \emph{monitorable} and \emph{non-monitorable}, respectively. In this paper, we revise the notion of monitorability from a practical perspective, where we show how \emph{non-monitorable} properties can still be used to generate \emph{partial} monitors, which can partially check the properties. Finally, we present the implications both from a theoretical and practical perspectives. 

\end{abstract}

\newcommand{\ev}{ev}
\newcommand{\univers}{\mathbb{U}}
\newcommand{\natSet}{\mathbb{N}}
\newcommand{\boolSet}{\mathbb{B}}
\newcommand{\presumablyTrue}{?_{\top}}
\newcommand{\presumablyFalse}{?_{\bot}}
\newcommand{\unknown}{?}
\newcommand{\giveup}{\upchi}
\newcommand{\cntxt}{\kappa}
\newcommand{\combine}{\chi}
\newcommand{\refine}{\otimes}
\newcommand{\aggregate}{\circ}
\newcommand{\multimodelSet}{\modelSet^{M}}
\newcommand{\compose}{\diamond}
\newcommand{\setting}{\mathcal{ST}}
\newcommand{\enrich}[1]{\mathsterling({#1})}
\newcommand{\decisionSet}{D}
\newcommand{\emptyTrace}{\epsilon}
\newcommand{\trace}{\sigma}
\newcommand{\continuation}{u}
\newcommand{\inftrace}{u}
\newcommand{\systemTraces}{tr(\eventSet)}
\newcommand{\spec}{\varphi}
\newcommand{\specSet}{\Phi}
\newcommand{\modelSet}{\Psi}
\newcommand{\functionSet}{\mathcal{F}}
\newcommand{\system}{S}
\newcommand{\eventSet}{\Sigma}
\newcommand{\predicate}{P}
\newcommand{\project}{\pi}
\newcommand{\sem}[1]{\llbracket {#1} \rrbracket}
\newcommand{\component}{\mathcal{C}}
\newcommand{\model}[1]{\ifx&#1&{\psi}\else \psi_{#1}\fi}
\newcommand{\function}[1]{\ifx&#1&{F}\else F_{#1}\fi}
\newcommand{\compMonitor}[1]{CoMon_{{#1}}}
\newcommand{\compMonitorAppl}[2]{CoMon_{{#1}}({#2})}
\newcommand{\decompose}{decomp}
\newcommand{\monitor}[3]{Mon_{{#1}, {#2}}^{#3}}
\newcommand{\choreographedMonitor}[3]{ChorMMon_{{#1}, {#2}}^{#3}}
\newcommand{\orchestratedMonitor}[3]{OrchMMon_{{#1}, {#2}}^{#3}}
\newcommand{\cenMonitor}[3]{CenMMon_{{#1}, {#2}}^{#3}}
\newcommand{\monitorAppl}[4]{Mon_{{#1}, {#2}}^{#3}({#4})}
\newcommand{\choreographedMonitorAppl}[4]{ChorMMon_{{#1}, {#2}}^{#3}({#4})}
\newcommand{\orchestratedMonitorAppl}[4]{OrchMMon_{{#1}, {#2}}^{#3}({#4})}
\newcommand{\cenMonitorAppl}[4]{CenMMon_{{#1}, {#2}}^{#3}({#4})}
\newcommand{\stmonitor}[1]{Mon_{{#1}}}
\newcommand{\stmonitorAppl}[2]{Mon_{{#1}}({#2})}
\newcommand{\modelEvents}[1]{\eventSet_{\model{#1}}}
\newcommand{\modelSetEvents}{\eventSet_\modelSet}
\newtheorem{observation}{\textsc{Observation}}
\newcommand{\modelTraces}[1]{tr(\model{#1})}
\newcommand{\modelSetTraces}{tr(\modelSet)}
\newcommand{\traces}[1]{tr({#1})}
\newcommand{\trans}[1]{\stackrel{{#1}}{\rightarrow}}
\newcommand{\combinedModel}{{\model{c}}}

\newcommand{\always}{\square}

\newtheorem{definition}{Definition}
\newtheorem{example}{Example}

\newcommand{\univmonitorable}{\forall_{PZ}\textrm{-}monitorable}
\newcommand{\existmonitorable}{\exists_{PZ}\textrm{-}monitorable}
\newcommand{\univmonitorableclass}{\forall_{PZ}}
\newcommand{\existmonitorableclass}{\exists_{PZ}}

\newcommand{\monitorable}[1]{{#1}\textrm{-}monitorable}

\newcommand{\agnote}[1]{{\todo[color=lightblue, inline]{Angelo: #1}}}
\newcommand{\rcnote}[1]{{\todo[color=red!65, inline]{Rafael: #1}}}

\newcommand{\until}{\ensuremath{\mathop{\mathrm{\mathbf{U}}}}}

\newcommand{\new}[1]{\textcolor{red}{#1}}

\section{Introduction}
Runtime Verification (RV) is a well-known lightweight formal verification technique~\cite{DBLP:series/lncs/BartocciFFR18}. Similar to other existing formal verification techniques, such as Model Checking~\cite{clarke1997model} and Theorem Proving~\cite{DBLP:books/lib/Loveland78}, it aims to verify the system behaviour, usually referred to as the System Under Analysis (SUA). Such a system can be composed of both software and hardware components, and the formal verification technique of choice is used to verify that everything works as expected. 

RV achieves the verification of the SUA through monitoring. Specifically, starting from a formal property expressed in some formalism of choice, one (or multiple) monitors are generated. A monitor is a device that, given a sequence of events (a trace) that are generated by the system execution, it verifies the conformance of such a trace with respect to the formal property. Since the trace can be generated at runtime, the monitor can inform the system's users about unexpected behaviours; \textit{i.e.}, events which violate the formal specification.

Differently from other formal verification techniques, RV is performed on the execution of the system. The formal properties are verified on traces of events generated by actual system executions. This is an important difference with respect to more traditional formal verification techniques such as Model Checking, where the verification is performed statically over an abstracted model of the system. RV does not require any model, nor any other information apart from execution traces, which makes it well suited to be used in \emph{black-box} scenarios where not much is known about the SUA, such as in autonomous systems. Moreover, from a computational perspective RV performs better than traditional verification techniques; since monitors only take as input what the SUA produces, without the need of going through a model. It has been shown that RV offers polynomial time behaviour with respect to the length of the analysed trace~\cite{DBLP:journals/jlp/LeuckerS09}.

Providing certification for reliable autonomous systems is not an easy task~\cite{DBLP:journals/aamas/FisherMRSWY21}. Formal verification of monolithic systems is already hard; to apply such techniques in the context of autonomous, cyber-physical, or even robotic systems makes it even more complicated. This is mainly due to the fact that these systems are intrinsically unpredictable~\cite{DBLP:journals/aim/LyonsCWS17}; especially when Machine Learning techniques are involved, such as Neural Networks~\cite{DBLP:books/lib/HertzKP91}. In these scenarios, RV may be of help, especially since it does not require a model of the system and it can be deployed at runtime while the system is still running. In this way, even though the system offers some unpredictable aspects, by adding monitors it is possible to improve the system reliability. In fact, since monitors can be deployed together with the system, the monitors will be there to detect and possibly react accordingly in case of unexpected behaviours (\textit{e.g.}, by triggering or implementing some mechanism to handle such failures).

Unfortunately, some formal properties cannot be monitored at runtime. A formal property is considered \emph{monitorable}, when it is possible to synthesise a monitor which verifies it. In turn, a formal property is denoted as \emph{non-monitorable}, when it is not possible to synthesise such a monitor. In literature, there exists different definitions on the requirements for a property to be considered monitorable~\cite{DBLP:journals/entcs/KimKLSV02}. Nonetheless, the most common requirement for a property to be monitorable is that it should always be possible to conclude the satisfaction or violation of the property. Which means that there should not exist traces of events which make the property never satisfied nor violated. The reason such kind of properties are usually avoided is that the resulting monitors might be useless, since they may never be able to conclude anything about the SUA. Because of this, RV approaches usually focus on monitorable fragments of the properties to analyse. However, as we will show in this paper, some of these non-monitorable properties may still be worth to be analysed at runtime, even though if only partially. Therefore, we pose the following research question:

\begin{center}
    \textit{Is it possible for monitors that are synthesised by non-monitorable properties to be used in practice?}
\end{center}

We suggest that a viable answer to this question is using \emph{partial monitoring}. Since a non-monitorable property does not give any assurance on satisfaction or violation, we need to synthesise monitors which are capable of \emph{giving up} on the verification process when it is clear that it will never arrive at any conclusion. This is especially relevant in the context of autonomous systems, where the availability of computational power and memory might be limited (\textit{e.g.}, in embedded systems). In such context, the presence of a component (the monitor) that arrives at a certain state where it does not do anything useful anymore is not only pointless, but it is also a waste of resources which could be better allocated. Thus, it is important to have monitors capable of giving up on the verification of a property, in order to safely reclaim otherwise used resources. At the same time, by using partial monitors we can be less restrictive on the kind of formal properties we are allowed to use, since a non-monitorable property can still be partially monitored at runtime.

As a proof of concept, in this work we exemplify our approach in a robotic application where an autonomous rover is deployed into a nuclear facility to perform remote inspection. In this scenario, the dynamic environment makes it hard to formally verify properties using traditional formal methods such as model checking. Thus, we can use RV to formally verify at runtime how the rover behaves. Using this application, we show that non-monitorable formal properties have parts of it that can still be verified using a partial monitor, as long as the monitor is capable of detecting when to give up. For example, we can have partial monitors to detect that the rover does not stay in areas with high-level of radiation, but if it observes a bad event (\textit{e.g.}, an event that would cause the monitor to be stuck in inconclusive states), which would render the monitor useless, then it needs to give up.

In this paper, we briefly revise the notion of \emph{monitorability}~\cite{DBLP:journals/entcs/KimKLSV02,DBLP:conf/sefm/AcetoAFIL19,DBLP:series/lncs/BartocciFFR18,DBLP:conf/fm/PnueliZ06}, where we focus more on its engineering implications for what concerns the monitor synthesis. To do so, we present a straightforward extension of the standard monitor synthesis for temporal properties in Linear Temporal Logic (LTL), where we take into consideration that a monitor could fail to completely verify an LTL property. We show how we can achieve this reasoning at the monitor level, instead of the more standard way of doing this at the property specification level. Specifically, we reduce the problem of a monitor recognising when to give up to a reachability problem inside the monitor representation.

The remainder of this paper is structured as follows.
Section~\ref{sec:preliminaries} presents some background definitions and notation that are used in the paper.
Section~\ref{sec:monitorability} revises the notion of monitorability and reports related works in literature.
Section~\ref{sec:partial-monitoring} proposes our contribution, where the notion of partial monitoring is introduced.
Section~\ref{sec:example} demonstrates the use of partial monitors in an autonomous rover performing remote inspection tasks.
In Section~\ref{sec:implementation}, we give the details on how the approach described in this paper has been implemented.
Section~\ref{sec:discussion} discusses the approach and its engineering implications in the monitor synthesis.
Finally, Section~\ref{sec:conclusions} concludes the paper with final remarks and future research directions.

\section{Preliminaries}
\label{sec:preliminaries}

A system $\system$ has an \textit{alphabet} $\eventSet$ containing all of its observable events. Given an alphabet $\eventSet$, a \emph{trace} $\trace=ev_0 ev_1 \ldots$, is a sequence of events in $\eventSet$. $\trace(i)$ is the i-th element of $\trace$ (\textit{i.e.}, $ev_i$), $\trace^i$ is the suffix of $\trace$ starting from $i$ (\textit{i.e.}, $ev_i ev_{i+1} \ldots$), $\eventSet^*$ is the \emph{set of all possible finite traces} over $\eventSet$, and $\eventSet^\omega$ is the \emph{set of all possible infinite traces} over $\eventSet$.

The standard formalism to specify formal properties in RV is propositional Linear Temporal Logic (LTL~\cite{DBLP:conf/focs/Pnueli77}). 
The relevant parts of the syntax of LTL are as follows:
{\small
$$
\spec = true \;|\; false \;|\; ev \;|\; (\spec\land\spec') \;|\; (\spec\lor\spec') \;|\; \lnot\spec \;|\; (\spec \;\until\; \spec') \;|\; \Circle\spec 
$$
}
where $ev\in\eventSet$ is an event (a proposition), $\spec$ is a formula, $\until$ stands for \emph{until}, and $\Circle$ stands for \emph{next-time}. In the rest of the paper, we also use the standard derived operators, such as $(\spec\rightarrow\spec')$ instead of $(\lnot\spec\lor\spec')$, $\spec \;R\; \spec'$ instead of $\lnot(\lnot\spec \until \lnot\spec')$, $\square \spec$ (\emph{always} $\spec$) instead of ($false \;R\; \spec$), and $\lozenge \spec$ (\emph{eventually} $\spec$) instead of ($true \until \spec$).

Let $\trace\in\eventSet^\omega$ be an infinite sequence of events over $\eventSet$, the semantics of LTL is as follows:
\begin{eqnarray*}
\trace &\models& ev \textrm{ if } ev \in \trace(0)\\
\trace &\models& \lnot\spec \textrm{ if } \trace \not\models \spec\\
\trace &\models& \spec\land\spec' \textrm{ if } \trace \models \spec \textrm{ and } \trace \models \spec'\\
\trace &\models& \spec\lor\spec' \textrm{ if } \trace \models \spec \textrm{ or } \trace \models \spec'\\
\trace &\models& \Circle\spec \textrm{ if } \trace^1\models\spec\\
\trace &\models& \spec  \until \spec' \textrm{ if } \exists_{i \geq 0}.\trace^i\models\spec' \textrm{ and } \forall_{0 \leq j < i}.\trace^j\models\spec
\end{eqnarray*}
A trace $\trace$ satisfies an atomic proposition ($ev$), if the event $ev$ belongs to the head (first element) of $\trace$; which means, $ev$ has been observed as initial event of the trace $\trace$. A trace $\trace$ satisfies the negation of the LTL property $\spec$, if $\trace$ does not satisfy $\spec$. A trace $\trace$ satisfies the conjunction of two LTL properties, if $\trace$ satisfies both properties. A trace $\trace$ satisfies the disjunction of two LTL properties, if $\trace$ satisfies at least one of them. 
A trace $\trace$ satisfies next-time $\spec$, if the suffix of $\trace$ starting in the next step ($\trace^1$) satisfies $\spec$. Finally, a trace $\trace$ satisfies $\spec\until\spec'$, if there exists a suffix of $\trace$ s.t. $\spec'$ is satisfied, and for all suffixes before it, $\spec$ holds. 

Thus, given an LTL property $\spec$, we denote $\sem{\spec}$ the language of the property, \textit{i.e.}, the set of traces which satisfy $\spec$; namely $\sem{\spec} = \{ \trace \;|\; \trace \models \spec\}$.

In Definition~\ref{rv-def}, we present a general and formalism-agnostic definition of a monitor. As mentioned before, a monitor is a function that, given a trace of events in input, returns a verdict which denotes the satisfaction (resp. violation) of a formal property over the trace.

\begin{definition}[Monitor]\label{rv-def}
Let~$\system$ be a system with alphabet~$\eventSet$,
and~$\spec$ be an LTL property. Then, a monitor for $\spec$ is a function
$\stmonitor{\spec}:\eventSet^*\rightarrow\mathbb{B}_3$, where
$\mathbb{B}_3=\{\top, \bot, \unknown \}$:
$$
\stmonitorAppl{\spec}{\trace} =
\left\{
\bgroup
\def\arraystretch{1.2}
  \begin{tabular}{cl}
  $\top$ & {\qquad$\forall_{\continuation \in \eventSet^\omega}.\trace \bullet \continuation \in \sem{\spec}$}\\
  $\bot$ & {\qquad$\forall_{\continuation \in \eventSet^\omega}.\trace \bullet \continuation \notin\sem{\spec}$}\\
  $\unknown$ & {\qquad$otherwise$}\\
  \end{tabular}
\egroup
\right.
$$
where $\bullet$ is the standard trace concatenation operator.
\end{definition}
Intuitively, a monitor returns~$\top$ if all continuations ($\continuation$) of $\trace$ satisfy $\spec$; $\bot$ if all possible
continuations of $\trace$ violate $\spec$; $\unknown$ otherwise. The first two outcomes are standard representations of satisfaction and violation, while the third is specific to RV. In more detail, it denotes when the monitor cannot conclude any verdict yet. This is closely related to the fact that RV is applied while the system is still running, and not all information about it are available. For instance, a property might be currently satisfied (resp. violated) by the system, but violated (resp. satisfied) in the (still unknown) future. The monitor can only safely conclude any of the two final verdicts ($\top$ or $\bot$) if it is sure such verdict will never change. The addition of the third outcome symbol $?$ helps the monitor to represent its position of uncertainty w.r.t. the current system execution.

A monitor function is usually implemented as a Finite State Machine (FSM), specifically a Moore machine (FSM where the output value of a state is only determined by the state)~\cite{DBLP:conf/fsttcs/BauerLS06,10.1145/2000799.2000800}. 
A Moore machine can be defined as a tuple $\langle Q,q_0,\Sigma,O,\delta,\gamma \rangle$, where $Q$ is a finite set of states, $q_0$ is the initial state, $\Sigma$ is the input alphabet, $O$ is the output alphabet, $\delta:Q\times\Sigma\rightarrow Q$ is the transition function mapping a state and an event to the next state, and $\gamma:Q\rightarrow O$ is the function mapping a state to the output alphabet.

\begin{figure}[h!]
\[
\hspace*{-0.5cm}
\xymatrix @C=0.3em @R=0.3em{ 
{Input} & {(i)Formula} & {(ii)NBA} & {(iii) \textrm{Emptiness per state}} & {(iv) NFA} & {(v) DFA} & {(vi) \textrm{Moore machine}} \\
& {\spec} \ar[r] & {A^{\spec}} \ar[r] & {F^{\spec}} \ar[r] & {{\hat{A}}^{\spec}} \ar[r] & {{\tilde{A}}^{\spec}} \ar[rd] & \\
{\spec} \ar[ru] \ar[rd] & & & & & & {\stmonitor{\spec}} \\
& {\lnot \spec} \ar[r] & {A^{\lnot \spec}} \ar[r] & {F^{\lnot \spec}} \ar[r] & {{\hat{A}}^{\lnot \spec}} \ar[r] & {{\tilde{A}}^{\lnot \spec}} \ar[ru] & \\
}
\]
\caption{Steps required to generate an FSM from an LTL formula $\spec$. NBA is Non-deterministic B{\"u}chi Automaton, NFA is Non-deterministic Finite Automaton, and DFA is Deterministic Finite Automaton.}\label{fsm-steps-fig}
\end{figure}
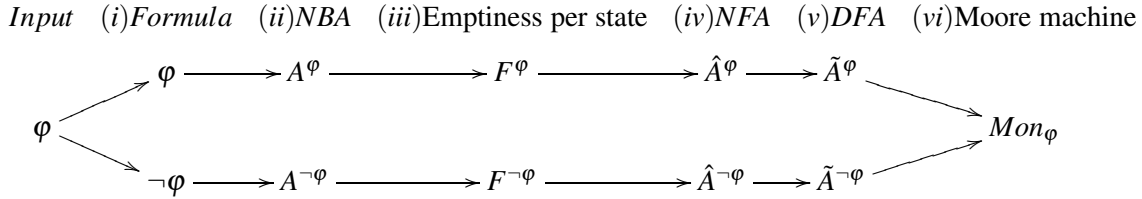

In~\cite{10.1145/2000799.2000800}, Bauer \textit{et al}. present the sequence of steps required to generate from an LTL formula $\spec$ the corresponding Moore machine instantiating the $\stmonitor{\spec}$ function (as summarised in Figure~\ref{fsm-steps-fig}).

Given an LTL property $\spec$, a series of transformations is performed on $\spec$, and its negation $\lnot\spec$. Considering $\spec$ in step \emph{(i)}, first, a corresponding B\"{u}chi Automaton $A^\spec$ is generated in step \emph{(ii)}. This can be obtained using Gerth \textit{et al}.'s algorithm ~\cite{DBLP:conf/pstv/GerthPVW95}. Such automaton recognises the set of infinite traces that satisfy $\spec$ (according to LTL semantics). Then, each state of $A^\spec$ is evaluated; the states that when selected as initial states in $A^\spec$ do not generate the empty language are then added to the $F^\spec$ set in step \emph{(iii)}. With such a set, a Non-deterministic Finite State Automaton $\hat{A}^\spec$ is obtained from $A^\spec$ by simply substituting the final states of $A^\spec$ with $F^\spec$ in step \emph{(iv)}. $\hat{A}^\spec$ recognises the finite traces (prefixes) that has at least one infinite continuation satisfying $\spec$ (since the prefix reaches a state in $F^\spec$). After that, $\hat{A}^\spec$ is transformed (Rabin–Scott powerset construction~\cite{DBLP:journals/ibmrd/RabinS59}) into its equivalent deterministic version $\tilde{A}^\spec$ in step \emph{(v)}; this is possible since deterministic and non-deterministic automata have the same expressive power. The exact same steps are performed on $\lnot\spec$, which bring to the generation of the $\tilde{A}^{\lnot\spec}$ counterpart. The difference between $\tilde{A}^\spec$ and $\tilde{A}^{\lnot\spec}$ is that the former recognises finite traces which have continuations satisfying $\spec$, while the latter recognises finite traces which have continuations violating $\spec$. Finally, a Moore machine can be generated as a standard automata product between $\tilde{A}^\spec$ and $\tilde{A}^{\lnot\spec}$ in the final step \emph{(vi)}, where the states are denoted as tuples $(q,q')$, with $q$ and $q'$ belonging to $\tilde{A}^\spec$ and $\tilde{A}^{\lnot\spec}$, respectively. The outputs are then determined as: $\top$ if $q'$ does not belong to the final states of $\tilde{A}^{\lnot\spec}$, $\bot$ if $q$ does not belong to the final states of $\tilde{A}^{\spec}$, and $\unknown$ otherwise.

\begin{example}\label{ex:fsm-example}
Let $\system$ be a system with alphabet $\eventSet=\{ev_1,ev_2,ev_3\}$, and $\spec = \lozenge ev_1$ be an LTL property to verify. In natural language, $\spec$ reads as: ``eventually event $ev_1$ is going to be observed''. The Moore machine implementing the monitor function $\stmonitor{\spec}$ is reported in Figure~\ref{fig:example-fsm}. As long as events $ev_2$ and $ev_3$ are observed, the Moore machine will stay in the initial state with output $\unknown$. As long as it stays in this state, there might be continuations where $ev_1$ will never be observed (\textit{i.e.}, the corresponding states in $\tilde{A}^\spec$ and $\tilde{A}^{\lnot\spec}$ are both finals). But, when $ev_1$ is observed, then the state changes to a positive state, with output $\top$. In fact, after observing $ev_1$, any trace determines positively $\spec$ since there is no continuation capable of violating $\spec$ (\textit{i.e.}, the corresponding state in $\tilde{A}^{\lnot\spec}$ is not final). 
\begin{figure}[ht]
\centering
\scalebox{0.8}{

\tikzset{every picture/.style={line width=0.75pt}} 

\begin{tikzpicture}[x=0.75pt,y=0.75pt,yscale=-1,xscale=1]

\draw  [fill={rgb, 255:red, 248; green, 231; blue, 28 }  ,fill opacity=0.7 ] (306,63) .. controls (306,49.19) and (317.19,38) .. (331,38) .. controls (344.81,38) and (356,49.19) .. (356,63) .. controls (356,76.81) and (344.81,88) .. (331,88) .. controls (317.19,88) and (306,76.81) .. (306,63) -- cycle ;
\draw  [fill={rgb, 255:red, 65; green, 117; blue, 5 }  ,fill opacity=0.7 ] (306,160) .. controls (306,146.19) and (317.19,135) .. (331,135) .. controls (344.81,135) and (356,146.19) .. (356,160) .. controls (356,173.81) and (344.81,185) .. (331,185) .. controls (317.19,185) and (306,173.81) .. (306,160) -- cycle ;
\draw    (331,88) .. controls (348.78,115.27) and (324.51,99.79) .. (330.48,132.4) ;
\draw [shift={(331,135)}, rotate = 257.84000000000003] [fill={rgb, 255:red, 0; green, 0; blue, 0 }  ][line width=0.08]  [draw opacity=0] (8.93,-4.29) -- (0,0) -- (8.93,4.29) -- cycle    ;
\draw    (331.23,17.97) -- (331.03,35) ;
\draw [shift={(331,38)}, rotate = 270.67] [fill={rgb, 255:red, 0; green, 0; blue, 0 }  ][line width=0.08]  [draw opacity=0] (8.93,-4.29) -- (0,0) -- (8.93,4.29) -- cycle    ;
\draw    (331,185) .. controls (320.39,222.4) and (405.62,186.09) .. (358.26,161.13) ;
\draw [shift={(356,160)}, rotate = 385.55] [fill={rgb, 255:red, 0; green, 0; blue, 0 }  ][line width=0.08]  [draw opacity=0] (8.93,-4.29) -- (0,0) -- (8.93,4.29) -- cycle    ;
\draw  [fill={rgb, 255:red, 248; green, 231; blue, 28 }  ,fill opacity=0.7 ] (442,181.5) .. controls (442,176.81) and (445.81,173) .. (450.5,173) .. controls (455.19,173) and (459,176.81) .. (459,181.5) .. controls (459,186.19) and (455.19,190) .. (450.5,190) .. controls (445.81,190) and (442,186.19) .. (442,181.5) -- cycle ;
\draw  [fill={rgb, 255:red, 65; green, 117; blue, 5 }  ,fill opacity=0.7 ] (442,203.5) .. controls (442,198.81) and (445.81,195) .. (450.5,195) .. controls (455.19,195) and (459,198.81) .. (459,203.5) .. controls (459,208.19) and (455.19,212) .. (450.5,212) .. controls (445.81,212) and (442,208.19) .. (442,203.5) -- cycle ;
\draw  [dash pattern={on 4.5pt off 4.5pt}] (426,145) -- (607,145) -- (607,236.83) -- (426,236.83) -- cycle ;
\draw    (450.23,147.97) -- (450.03,165) ;
\draw [shift={(450,168)}, rotate = 270.67] [fill={rgb, 255:red, 0; green, 0; blue, 0 }  ][line width=0.08]  [draw opacity=0] (8.93,-4.29) -- (0,0) -- (8.93,4.29) -- cycle    ;
\draw    (341,86) .. controls (374.55,103.48) and (397.31,36.6) .. (351.15,45.21) ;
\draw [shift={(348.23,45.83)}, rotate = 346.5] [fill={rgb, 255:red, 0; green, 0; blue, 0 }  ][line width=0.08]  [draw opacity=0] (8.93,-4.29) -- (0,0) -- (8.93,4.29) -- cycle    ;

\draw (330,63) node  [font=\large] [align=left] {?};
\draw (331,160) node  [font=\large] [align=left] {$\displaystyle \top $};
\draw (316.27,107.31) node   [align=left] {$\displaystyle ev_{1}$};
\draw (408.28,64.31) node   [align=left] {$\displaystyle ev_{2} ,ev_{3}$};
\draw (369.28,201.31) node   [align=left] {$\displaystyle *$};
\draw (542.28,181.31) node   [align=left] {inconclusive state};
\draw (542.28,203.31) node   [align=left] {positive state};
\draw (452.28,224.31) node   [align=left] {$\displaystyle *$};
\draw (542.28,223.31) node   [align=left] {any event};
\draw (542.28,159.31) node   [align=left] {initial state};

\end{tikzpicture}

}

\caption{Moore machine instantiated for the monitor generated by $\spec$ of Example~\ref{ex:fsm-example}.}
\label{fig:example-fsm}
\end{figure}
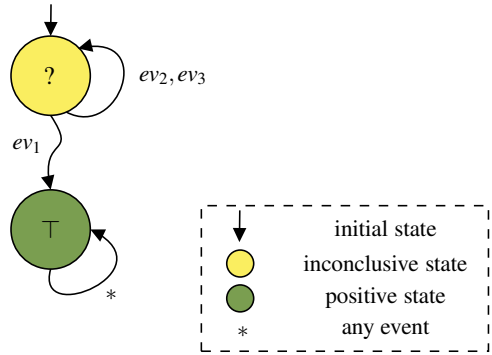
\end{example}

\section{Monitorability}
\label{sec:monitorability}

\emph{Monitorability}~\cite{DBLP:journals/entcs/KimKLSV02} refers to the branch of RV focused on the delineation of which formal properties can be monitored. It is crucial to understand monitorability for performing efficient verification  at runtime of formal properties. However, the level of detail such notion is defined in the literature may vary. It has a wide range of definitions, some are more restrictive, while others are more flexible. 
A thorough presentation of the existing variations of monitorability can be found in~\cite{DBLP:conf/sefm/AcetoAFIL19}; where the authors report a complete guide on monitorability and its uses.

In this paper, we consider the definitions of monitorability introduced by Pnueli and Zaks~\cite{DBLP:conf/fm/PnueliZ06}, where the concept of monitorability was generalised w.r.t. its first appearance~\cite{DBLP:journals/entcs/KimKLSV02}. We chose their view of monitoriability since it is one of the most commonly cited by the community and it is less restrictive on the set of non-monitorable properties than other definitions found in the literature.

\begin{definition}\label{def:monitorable}
A property $\spec$ is $\monitorable{\trace}$, where $\trace\in\eventSet^*$, if there is some $\continuation\in\eventSet^*$ such that $\spec$ is positively or negatively determined by $\trace\bullet\continuation$.
\end{definition}

Definition~\ref{def:monitorable} states that a property $\spec$ is considered \emph{monitorable} with respect to a finite trace of events $\trace$, if we can find at least one trace $\continuation$, such that $\spec$ is satisfied (\textit{i.e.}, $\trace$ is a \emph{good} prefix) or violated (\textit{i.e.}, $\trace$ is a \emph{bad} prefix) by the resulting concatenated trace $\trace\bullet\continuation$. Intuitively, if a property is $\monitorable{\trace}$, we know that for at least one possible trace of events the monitor will be able to conclude the satisfaction or violation of $\spec$.

Following this reasoning, we define four different notions of monitorable property. 
We start from less restrictive and move towards more restrictive notions. Definition~\ref{def:existentially-monitorable} uses a more relaxed notion of monitorability, where a property $\spec$ is considered \emph{existentially} monitorable when for at least one trace of events $\trace\in\eventSet^*$ it is possible to find a continuation for which $\spec$ is either satisfied or violated. Intuitively, this means that some trace can bring the monitor to never conclude the satisfaction or violation of $\spec$. In the literature, these properties are also known as \emph{weak monitorable}~\cite{DBLP:conf/icse/ChenWWS18}.

\begin{definition}\label{def:existentially-monitorable}
A property $\spec$ is (existentially Pnueli-Zaks) $\existmonitorable$ if it is $\monitorable{\trace}$ for some finite trace $\trace\in\eventSet^*$. The class of all $\existmonitorable$ properties is denoted as $\existmonitorableclass$.
\end{definition}

\begin{figure}[ht]
\centering
\scalebox{0.8}{

\tikzset{every picture/.style={line width=0.75pt}} 

\begin{tikzpicture}[x=0.75pt,y=0.75pt,yscale=-1,xscale=1]

\draw  [fill={rgb, 255:red, 248; green, 231; blue, 28 }  ,fill opacity=0.7 ] (306,63) .. controls (306,49.19) and (317.19,38) .. (331,38) .. controls (344.81,38) and (356,49.19) .. (356,63) .. controls (356,76.81) and (344.81,88) .. (331,88) .. controls (317.19,88) and (306,76.81) .. (306,63) -- cycle ;
\draw  [fill={rgb, 255:red, 248; green, 231; blue, 28 }  ,fill opacity=0.7 ] (404,150) .. controls (404,136.19) and (415.19,125) .. (429,125) .. controls (442.81,125) and (454,136.19) .. (454,150) .. controls (454,163.81) and (442.81,175) .. (429,175) .. controls (415.19,175) and (404,163.81) .. (404,150) -- cycle ;
\draw  [fill={rgb, 255:red, 248; green, 231; blue, 28 }  ,fill opacity=0.7 ] (187,146) .. controls (187,132.19) and (198.19,121) .. (212,121) .. controls (225.81,121) and (237,132.19) .. (237,146) .. controls (237,159.81) and (225.81,171) .. (212,171) .. controls (198.19,171) and (187,159.81) .. (187,146) -- cycle ;
\draw  [fill={rgb, 255:red, 208; green, 2; blue, 27 }  ,fill opacity=0.7 ] (294,152) .. controls (294,138.19) and (305.19,127) .. (319,127) .. controls (332.81,127) and (344,138.19) .. (344,152) .. controls (344,165.81) and (332.81,177) .. (319,177) .. controls (305.19,177) and (294,165.81) .. (294,152) -- cycle ;
\draw  [fill={rgb, 255:red, 65; green, 117; blue, 5 }  ,fill opacity=0.7 ] (187,243) .. controls (187,229.19) and (198.19,218) .. (212,218) .. controls (225.81,218) and (237,229.19) .. (237,243) .. controls (237,256.81) and (225.81,268) .. (212,268) .. controls (198.19,268) and (187,256.81) .. (187,243) -- cycle ;
\draw    (356,63) .. controls (435.61,62.97) and (419.91,75.45) .. (428.45,122.1) ;
\draw [shift={(429,125)}, rotate = 258.74] [fill={rgb, 255:red, 0; green, 0; blue, 0 }  ][line width=0.08]  [draw opacity=0] (8.93,-4.29) -- (0,0) -- (8.93,4.29) -- cycle    ;
\draw    (429,175) .. controls (418.39,212.4) and (503.62,176.09) .. (456.26,151.13) ;
\draw [shift={(454,150)}, rotate = 385.55] [fill={rgb, 255:red, 0; green, 0; blue, 0 }  ][line width=0.08]  [draw opacity=0] (8.93,-4.29) -- (0,0) -- (8.93,4.29) -- cycle    ;
\draw    (331,88) .. controls (348.78,115.27) and (313.1,92.18) .. (318.5,124.41) ;
\draw [shift={(319,127)}, rotate = 257.84000000000003] [fill={rgb, 255:red, 0; green, 0; blue, 0 }  ][line width=0.08]  [draw opacity=0] (8.93,-4.29) -- (0,0) -- (8.93,4.29) -- cycle    ;
\draw    (319,177) .. controls (308.39,214.4) and (393.62,178.09) .. (346.26,153.13) ;
\draw [shift={(344,152)}, rotate = 385.55] [fill={rgb, 255:red, 0; green, 0; blue, 0 }  ][line width=0.08]  [draw opacity=0] (8.93,-4.29) -- (0,0) -- (8.93,4.29) -- cycle    ;
\draw    (306,63) .. controls (211.67,59.03) and (224.81,69.64) .. (212.58,118.73) ;
\draw [shift={(212,121)}, rotate = 284.54] [fill={rgb, 255:red, 0; green, 0; blue, 0 }  ][line width=0.08]  [draw opacity=0] (8.93,-4.29) -- (0,0) -- (8.93,4.29) -- cycle    ;
\draw    (212,171) .. controls (229.78,198.27) and (205.51,182.79) .. (211.48,215.4) ;
\draw [shift={(212,218)}, rotate = 257.84000000000003] [fill={rgb, 255:red, 0; green, 0; blue, 0 }  ][line width=0.08]  [draw opacity=0] (8.93,-4.29) -- (0,0) -- (8.93,4.29) -- cycle    ;
\draw    (331.23,17.97) -- (331.03,35) ;
\draw [shift={(331,38)}, rotate = 270.67] [fill={rgb, 255:red, 0; green, 0; blue, 0 }  ][line width=0.08]  [draw opacity=0] (8.93,-4.29) -- (0,0) -- (8.93,4.29) -- cycle    ;
\draw    (224,167) .. controls (213.39,204.4) and (286.97,171.97) .. (239.26,147.13) ;
\draw [shift={(237,146)}, rotate = 385.55] [fill={rgb, 255:red, 0; green, 0; blue, 0 }  ][line width=0.08]  [draw opacity=0] (8.93,-4.29) -- (0,0) -- (8.93,4.29) -- cycle    ;
\draw    (212,268) .. controls (201.39,305.4) and (286.62,269.09) .. (239.26,244.13) ;
\draw [shift={(237,243)}, rotate = 385.55] [fill={rgb, 255:red, 0; green, 0; blue, 0 }  ][line width=0.08]  [draw opacity=0] (8.93,-4.29) -- (0,0) -- (8.93,4.29) -- cycle    ;
\draw  [fill={rgb, 255:red, 248; green, 231; blue, 28 }  ,fill opacity=0.7 ] (518,197.5) .. controls (518,192.81) and (521.81,189) .. (526.5,189) .. controls (531.19,189) and (535,192.81) .. (535,197.5) .. controls (535,202.19) and (531.19,206) .. (526.5,206) .. controls (521.81,206) and (518,202.19) .. (518,197.5) -- cycle ;
\draw  [fill={rgb, 255:red, 208; green, 2; blue, 27 }  ,fill opacity=0.7 ] (518,217.5) .. controls (518,212.81) and (521.81,209) .. (526.5,209) .. controls (531.19,209) and (535,212.81) .. (535,217.5) .. controls (535,222.19) and (531.19,226) .. (526.5,226) .. controls (521.81,226) and (518,222.19) .. (518,217.5) -- cycle ;
\draw  [fill={rgb, 255:red, 65; green, 117; blue, 5 }  ,fill opacity=0.7 ] (518,238.5) .. controls (518,233.81) and (521.81,230) .. (526.5,230) .. controls (531.19,230) and (535,233.81) .. (535,238.5) .. controls (535,243.19) and (531.19,247) .. (526.5,247) .. controls (521.81,247) and (518,243.19) .. (518,238.5) -- cycle ;
\draw  [dash pattern={on 4.5pt off 4.5pt}] (502,161) -- (683,161) -- (683,291) -- (502,291) -- cycle ;
\draw    (526.23,163.97) -- (526.03,181) ;
\draw [shift={(526,184)}, rotate = 270.67] [fill={rgb, 255:red, 0; green, 0; blue, 0 }  ][line width=0.08]  [draw opacity=0] (8.93,-4.29) -- (0,0) -- (8.93,4.29) -- cycle    ;

\draw (330,63) node  [font=\large] [align=left] {?};
\draw (429,150) node  [font=\large] [align=left] {?};
\draw (212,146) node  [font=\large] [align=left] {?};
\draw (319,152) node  [font=\large] [align=left] {$\displaystyle \bot $};
\draw (212,243) node  [font=\large] [align=left] {$\displaystyle \top $};
\draw (269.27,50.31) node   [align=left] {$\displaystyle ev_{1}$};
\draw (197.27,190.31) node   [align=left] {$\displaystyle ev_{2}$};
\draw (262.27,190.31) node   [align=left] {$\displaystyle *\setminus ev_{2}$};
\draw (351.28,195.31) node   [align=left] {$\displaystyle *$};
\draw (464.28,192.31) node   [align=left] {$\displaystyle *$};
\draw (408.28,52.31) node   [align=left] {$\displaystyle ev_{3}$};
\draw (295.27,94.31) node   [align=left] {$\displaystyle ev_{2} ,ev_{4}$};
\draw (250.27,284.31) node   [align=left] {$\displaystyle *$};
\draw (618.28,197.31) node   [align=left] {inconclusive state};
\draw (618.28,217.31) node   [align=left] {negative state};
\draw (618.28,238.31) node   [align=left] {positive state};
\draw (528.28,259.31) node   [align=left] {$\displaystyle *$};
\draw (618.28,258.31) node   [align=left] {any event};
\draw (528.28,276) node   [align=left] {$\displaystyle *\setminus ev_{i}$};
\draw (618.28,275) node   [align=left] {any event but $\displaystyle ev_{i}$};
\draw (618.28,175.31) node   [align=left] {initial state};

\end{tikzpicture}

}

\caption{Moore machine of the $\existmonitorable$ property $\spec$ presented in Example~\ref{ex:existentiallyPZ}.}
\label{fig:existentiallyPZ}
\end{figure}
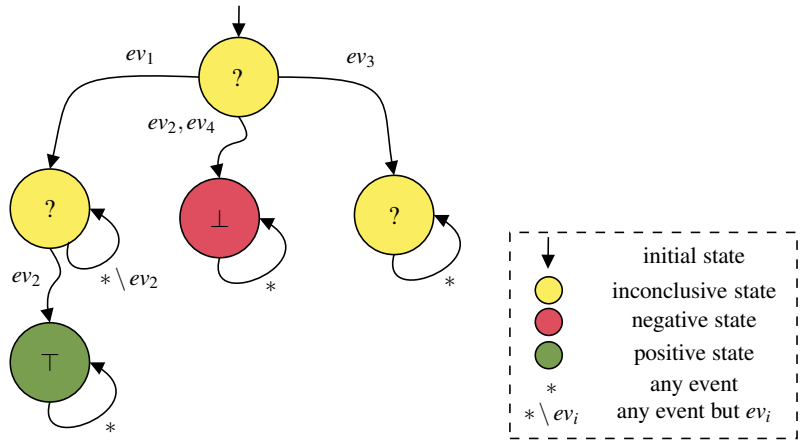

\begin{example}\label{ex:existentiallyPZ}
Let us assume $\spec=(ev_1 \land \lozenge ev_2)\lor(ev_3 \land \square\lozenge ev_4)$, and $\eventSet=\{ev_1,ev_2,ev_3,ev_4\}$. This is an example of a $\existmonitorable$ property, since we can find some $\trace\in\eventSet^*$ for which $\spec$ is $\monitorable{\trace}$; \textit{e.g.}, any trace starting with $ev_1$ can eventually satisfy $\spec$ (by observing $ev_2$). Furthermore, every trace $\trace\in\eventSet^*$ starting with $ev_3$ is not $\monitorable{\trace}$, since there is no continuation $\continuation\in\eventSet^*$ s.t. $\trace\bullet\continuation$ positively or negatively determines $\spec$.
Figure~\ref{fig:existentiallyPZ} reports the monitor obtained by $\spec$. 
\end{example}

Definition~\ref{def:universally-monitorable} introduces a more restrictive notion of monitorability, where a property $\spec$ is considered \emph{universally} monitorable when for all finite traces $\trace\in\eventSet^*$, it is possible to find a continuation for which $\spec$ is either satisfied or violated. This means that no trace can bring the monitor to never conclude the satisfaction or violation of $\spec$.

\begin{definition}\label{def:universally-monitorable}
A property $\spec$ is (universally Pnueli-Zaks) $\univmonitorable$ if it is $\monitorable{\trace}$ for all finite trace $\trace\in\eventSet^*$. The class of all $\univmonitorable$ properties is denoted as $\univmonitorableclass$.
\end{definition}

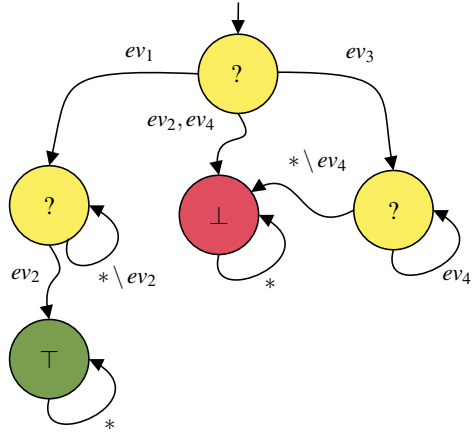
\begin{figure}[ht]
\centering
\scalebox{0.8}{

\tikzset{every picture/.style={line width=0.75pt}} 

\begin{tikzpicture}[x=0.75pt,y=0.75pt,yscale=-1,xscale=1]

\draw  [fill={rgb, 255:red, 248; green, 231; blue, 28 }  ,fill opacity=0.7 ] (306,63) .. controls (306,49.19) and (317.19,38) .. (331,38) .. controls (344.81,38) and (356,49.19) .. (356,63) .. controls (356,76.81) and (344.81,88) .. (331,88) .. controls (317.19,88) and (306,76.81) .. (306,63) -- cycle ;
\draw  [fill={rgb, 255:red, 248; green, 231; blue, 28 }  ,fill opacity=0.7 ] (187,146) .. controls (187,132.19) and (198.19,121) .. (212,121) .. controls (225.81,121) and (237,132.19) .. (237,146) .. controls (237,159.81) and (225.81,171) .. (212,171) .. controls (198.19,171) and (187,159.81) .. (187,146) -- cycle ;
\draw  [fill={rgb, 255:red, 208; green, 2; blue, 27 }  ,fill opacity=0.7 ] (294,152) .. controls (294,138.19) and (305.19,127) .. (319,127) .. controls (332.81,127) and (344,138.19) .. (344,152) .. controls (344,165.81) and (332.81,177) .. (319,177) .. controls (305.19,177) and (294,165.81) .. (294,152) -- cycle ;
\draw  [fill={rgb, 255:red, 65; green, 117; blue, 5 }  ,fill opacity=0.7 ] (187,243) .. controls (187,229.19) and (198.19,218) .. (212,218) .. controls (225.81,218) and (237,229.19) .. (237,243) .. controls (237,256.81) and (225.81,268) .. (212,268) .. controls (198.19,268) and (187,256.81) .. (187,243) -- cycle ;
\draw    (331,88) .. controls (348.78,115.27) and (313.1,92.18) .. (318.5,124.41) ;
\draw [shift={(319,127)}, rotate = 257.84000000000003] [fill={rgb, 255:red, 0; green, 0; blue, 0 }  ][line width=0.08]  [draw opacity=0] (8.93,-4.29) -- (0,0) -- (8.93,4.29) -- cycle    ;
\draw    (319,177) .. controls (308.39,214.4) and (393.62,178.09) .. (346.26,153.13) ;
\draw [shift={(344,152)}, rotate = 385.55] [fill={rgb, 255:red, 0; green, 0; blue, 0 }  ][line width=0.08]  [draw opacity=0] (8.93,-4.29) -- (0,0) -- (8.93,4.29) -- cycle    ;
\draw    (306,63) .. controls (211.67,59.03) and (224.81,69.64) .. (212.58,118.73) ;
\draw [shift={(212,121)}, rotate = 284.54] [fill={rgb, 255:red, 0; green, 0; blue, 0 }  ][line width=0.08]  [draw opacity=0] (8.93,-4.29) -- (0,0) -- (8.93,4.29) -- cycle    ;
\draw    (212,171) .. controls (229.78,198.27) and (205.51,182.79) .. (211.48,215.4) ;
\draw [shift={(212,218)}, rotate = 257.84000000000003] [fill={rgb, 255:red, 0; green, 0; blue, 0 }  ][line width=0.08]  [draw opacity=0] (8.93,-4.29) -- (0,0) -- (8.93,4.29) -- cycle    ;
\draw    (331.23,17.97) -- (331.03,35) ;
\draw [shift={(331,38)}, rotate = 270.67] [fill={rgb, 255:red, 0; green, 0; blue, 0 }  ][line width=0.08]  [draw opacity=0] (8.93,-4.29) -- (0,0) -- (8.93,4.29) -- cycle    ;
\draw    (224,167) .. controls (213.39,204.4) and (286.97,171.97) .. (239.26,147.13) ;
\draw [shift={(237,146)}, rotate = 385.55] [fill={rgb, 255:red, 0; green, 0; blue, 0 }  ][line width=0.08]  [draw opacity=0] (8.93,-4.29) -- (0,0) -- (8.93,4.29) -- cycle    ;
\draw    (212,268) .. controls (201.39,305.4) and (286.62,269.09) .. (239.26,244.13) ;
\draw [shift={(237,243)}, rotate = 385.55] [fill={rgb, 255:red, 0; green, 0; blue, 0 }  ][line width=0.08]  [draw opacity=0] (8.93,-4.29) -- (0,0) -- (8.93,4.29) -- cycle    ;
\draw  [fill={rgb, 255:red, 248; green, 231; blue, 28 }  ,fill opacity=0.7 ] (404,149) .. controls (404,135.19) and (415.19,124) .. (429,124) .. controls (442.81,124) and (454,135.19) .. (454,149) .. controls (454,162.81) and (442.81,174) .. (429,174) .. controls (415.19,174) and (404,162.81) .. (404,149) -- cycle ;
\draw    (356,62) .. controls (435.61,61.97) and (419.91,74.45) .. (428.45,121.1) ;
\draw [shift={(429,124)}, rotate = 258.74] [fill={rgb, 255:red, 0; green, 0; blue, 0 }  ][line width=0.08]  [draw opacity=0] (8.93,-4.29) -- (0,0) -- (8.93,4.29) -- cycle    ;
\draw    (429,174) .. controls (418.39,211.4) and (503.62,175.09) .. (456.26,150.13) ;
\draw [shift={(454,149)}, rotate = 385.55] [fill={rgb, 255:red, 0; green, 0; blue, 0 }  ][line width=0.08]  [draw opacity=0] (8.93,-4.29) -- (0,0) -- (8.93,4.29) -- cycle    ;
\draw    (404,149) .. controls (369.13,168.14) and (372.06,118.24) .. (341.64,136.12) ;
\draw [shift={(339.23,137.63)}, rotate = 326.31] [fill={rgb, 255:red, 0; green, 0; blue, 0 }  ][line width=0.08]  [draw opacity=0] (8.93,-4.29) -- (0,0) -- (8.93,4.29) -- cycle    ;

\draw (330,63) node  [font=\large] [align=left] {?};
\draw (212,146) node  [font=\large] [align=left] {?};
\draw (319,152) node  [font=\large] [align=left] {$\displaystyle \bot $};
\draw (212,243) node  [font=\large] [align=left] {$\displaystyle \top $};
\draw (269.27,50.31) node   [align=left] {$\displaystyle ev_{1}$};
\draw (197.27,190.31) node   [align=left] {$\displaystyle ev_{2}$};
\draw (261.27,191.31) node   [align=left] {$\displaystyle *\setminus ev_{2}$};
\draw (351.28,195.31) node   [align=left] {$\displaystyle *$};
\draw (295.27,94.31) node   [align=left] {$\displaystyle ev_{2} ,ev_{4}$};
\draw (250.27,284.31) node   [align=left] {$\displaystyle *$};
\draw (429,149) node  [font=\large] [align=left] {?};
\draw (469.28,191.31) node   [align=left] {$\displaystyle ev_{4}$};
\draw (408.28,51.31) node   [align=left] {$\displaystyle ev_{3}$};
\draw (381.28,118.31) node   [align=left] {$\displaystyle *\setminus ev_{4}$};

\end{tikzpicture}

}

\caption{Monitor (as Moore machine) of the $\univmonitorable$ property $\spec$ presented in Example~\ref{ex:universallyPZ}.}
\label{fig:universallyPZ}
\end{figure}

\begin{example}\label{ex:universallyPZ}
Let us assume $\spec=(ev_1 \rightarrow \lozenge ev_2)\lor(ev_3 \rightarrow \square ev_4)$, and $\eventSet=\{ev_1,ev_2,ev_3,ev_4\}$. This is an example of a $\univmonitorable$ property, since for every $\trace\in\eventSet^*$, the property is $\monitorable{\trace}$; we can always find a continuation $\continuation\in\eventSet^*$ s.t. the $\spec$ is positively or negatively determined. This can be seen on the left branch where we have the possibility to satisfy the property by observing (eventually) $ev_2$ (positive); while on the right branch, we have the possibility to violate the property by observing something different from $ev_4$ (negative). Figure~\ref{fig:universallyPZ} reports the monitor obtained by $\spec$.
\end{example}

Monitorable properties are defined according to Definition~\ref{def:universally-monitorable} in a variety of past research~\cite{10.1145/2000799.2000800,DBLP:series/lncs/BartocciFFR18,DBLP:journals/jlp/LeuckerS09,DBLP:conf/rv/HenzingerS20,DBLP:journals/entcs/KimKLSV02}. The reason for this is that any other notion of monitorability, such as the one proposed in Definition~\ref{def:existentially-monitorable}, does not give any guarantees on the monitor used to verify the property. Indeed, if we consider Definition~\ref{def:existentially-monitorable}, there is no guarantee that eventually the monitor will encounter a trace of events $\trace$ for which no continuation determines $\spec$ positively or negatively. If this happens, then the monitor will just become pointless, because it will remain in an inconclusive state forever (\textit{i.e.}, it will never conclude anything about $\spec$). In such scenarios, we follow the notation used in~\cite{10.1145/2000799.2000800} to refer to such a trace of events $\trace$ as an \emph{ugly} prefix, since it represents a case where nothing can (and nothing will) be concluded. In order to avoid these scenarios, more restrictive rules over monitorability are usually imposed; of which Definition~\ref{def:universally-monitorable} is a key example.

Next, we show that by restricting even more the notion of monitorability, we find \emph{Safety} and \emph{Co-Safety} properties~\cite{DBLP:journals/dc/AlpernS87}. Definition~\ref{def:safety} denotes the properties of the kind ``\textit{nothing bad will ever happen}''.

\begin{definition}\label{def:safety}
A property $\spec$ is a \emph{safety} property if every $\trace\not\in\sem{\spec}$ has a prefix that determines $\spec$ negatively. The class of safety properties is denoted as Safe.
\end{definition}

These properties can only be violated at runtime, which means the resulting monitor can only report negative and inconclusive verdicts. This is due to the fact that safety properties are satisfied only by infinite traces of events, and at runtime we only have access to finite traces.

An example of a safety property can be found in the right branch of $\spec$ in Example~\ref{ex:universallyPZ}, \textit{i.e.}, $\square ev_4$. This is a safety property where the expected behaviour is to observe $ev_4$ indefinitely. Thus, any $\trace\in\eventSet^*$ (the $\eventSet$ of Example~\ref{ex:universallyPZ}) can be extended with a continuation $\continuation\in\eventSet^*$ s.t. $\trace\bullet\continuation$ negatively determines $\spec$. There is no continuation $\continuation\in\eventSet^*$ s.t. $\trace\bullet\continuation$ positively determines 
$\spec$, but this is not actually required for being monitorable.

On the same level of restrictiveness, we have \emph{Co-Safety} properties. Definition~\ref{def:cosafety} denotes the properties of the kind ``something good will eventually happen''.

\begin{definition}\label{def:cosafety}
A property $\spec$ is a \emph{co-safety} property if every $\trace\in\sem{\spec}$ has a prefix that determines $\spec$ positively. The class of co-safety properties is denoted as CoSafe.
\end{definition}

These properties can only be satisfied at runtime, which means the resulting monitor can only report positive and inconclusive verdicts. This is due to the fact that co-safety properties are violated only by infinite traces of events.

An example of a co-safety property can be found in the left branch of $\spec$ in Example~\ref{ex:universallyPZ}, \textit{i.e.}, $\lozenge ev_2$. This is a co-safety property where the expected behaviour is to observe eventually $ev_2$. Thus, any $\trace\in\eventSet^*$ (the $\eventSet$ of Example~\ref{ex:universallyPZ}) can be extended with a continuation $\continuation\in\eventSet^*$ s.t. $\trace\bullet\continuation$ positively determines $\spec$. There is no continuation $\continuation\in\eventSet^*$ s.t. $\trace\bullet\continuation$ negatively determines 
$\spec$, but once again this is not required for being monitorable.

Note that, to check if a property belongs to the Safe class it is a PSPACE problem, while to check if a property is in the CoSafe class it is an EXPSPACE problem~\cite{DBLP:journals/fac/Sistla94}.

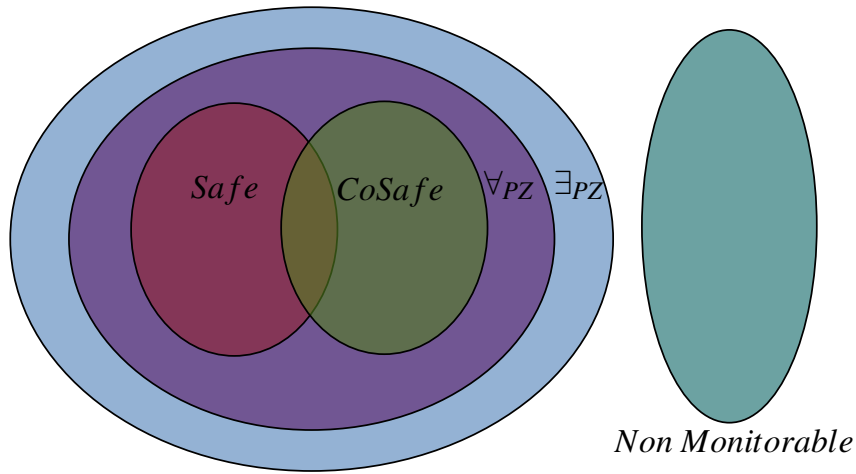
\begin{figure}[ht]
\centering
 \scalebox{0.88}{

\tikzset{every picture/.style={line width=0.75pt}} 

\begin{tikzpicture}[x=0.75pt,y=0.75pt,yscale=-1,xscale=1]

\draw  [fill={rgb, 255:red, 78; green, 127; blue, 184 }  ,fill opacity=0.6 ] (101,142.7) .. controls (101,69.25) and (178.35,9.7) .. (273.76,9.7) .. controls (369.17,9.7) and (446.52,69.25) .. (446.52,142.7) .. controls (446.52,216.15) and (369.17,275.7) .. (273.76,275.7) .. controls (178.35,275.7) and (101,216.15) .. (101,142.7) -- cycle ;
\draw  [fill={rgb, 255:red, 91; green, 26; blue, 105 }  ,fill opacity=0.6 ] (134.63,142.7) .. controls (134.63,82.22) and (196.92,33.2) .. (273.76,33.2) .. controls (350.6,33.2) and (412.89,82.22) .. (412.89,142.7) .. controls (412.89,203.18) and (350.6,252.2) .. (273.76,252.2) .. controls (196.92,252.2) and (134.63,203.18) .. (134.63,142.7) -- cycle ;
\draw  [fill={rgb, 255:red, 145; green, 33; blue, 47 }  ,fill opacity=0.6 ] (170.25,137.2) .. controls (170.25,97.16) and (196.72,64.7) .. (229.38,64.7) .. controls (262.04,64.7) and (288.52,97.16) .. (288.52,137.2) .. controls (288.52,177.24) and (262.04,209.7) .. (229.38,209.7) .. controls (196.72,209.7) and (170.25,177.24) .. (170.25,137.2) -- cycle ;
\draw  [fill={rgb, 255:red, 82; green, 127; blue, 31 }  ,fill opacity=0.6 ] (256.25,136.2) .. controls (256.25,96.16) and (282.72,63.7) .. (315.38,63.7) .. controls (348.04,63.7) and (374.52,96.16) .. (374.52,136.2) .. controls (374.52,176.24) and (348.04,208.7) .. (315.38,208.7) .. controls (282.72,208.7) and (256.25,176.24) .. (256.25,136.2) -- cycle ;
\draw  [fill={rgb, 255:red, 10; green, 100; blue, 100 }  ,fill opacity=0.6 ] (463,135.35) .. controls (463,73.14) and (485.44,22.7) .. (513.12,22.7) .. controls (540.8,22.7) and (563.23,73.14) .. (563.23,135.35) .. controls (563.23,197.56) and (540.8,248) .. (513.12,248) .. controls (485.44,248) and (463,197.56) .. (463,135.35) -- cycle ;

\draw (411,102) node [anchor=north west][inner sep=0.75pt]  [font=\Large] [align=left] {$\displaystyle \exists _{PZ}$};
\draw (371,102) node [anchor=north west][inner sep=0.75pt]  [font=\Large] [align=left] {$\displaystyle \forall _{PZ}$};
\draw (287,106) node [anchor=north west][inner sep=0.75pt]  [font=\Large] [align=left] {$\displaystyle CoSafe$};
\draw (203,105) node [anchor=north west][inner sep=0.75pt]  [font=\Large] [align=left] {$\displaystyle Safe$};
\draw (446,250) node [anchor=north west][inner sep=0.75pt]  [font=\Large] [align=left] {$\displaystyle Non\ Monitorable$};

\end{tikzpicture}

 }

\caption{Hierarchy of classes of monitorable properties.}
\label{fig:hierarchy}
\end{figure}

Figure~\ref{fig:hierarchy} represents the different monitorability classes as sets. On the left, the largest set corresponds to $\existmonitorableclass$, which only requires the properties to have at least one good prefix. Then, we have $\univmonitorableclass$, which requires the properties to have only good prefixes. After that, we find the \emph{Safe} and \emph{CoSafe} classes, which are included in $\univmonitorableclass$ by construction. Since for every safety (resp. co-safety) property and $\trace\in\eventSet^*$, we may find $\continuation\in\eventSet^*$ s.t. $\trace\bullet\continuation$ negatively (resp. positively) determines the property. On the right, we have the rest of the properties, which are considered \emph{non-monitorable}. These are the properties for which there is no trace $\trace\in\eventSet^*$ which determines the property neither positively nor negatively. Thus, properties for which all traces are ugly.

Note that the more restrictive the conditions for a property to be considered monitorable are, the less these properties will be used in practice for achieving RV. Thus, one has to find a good balance to be able to discard as few properties as possible, and at the same time to correctly handle the possible lack of guarantees on the resulting monitoring process. The presence of a monitor that can reach a state with nothing to report should be avoided, or at least detected and handled. 

In~\cite{DBLP:conf/fm/PnueliZ06}, Pnueli and Zaks propose a way to decide, given a finite trace $\trace$, if a property under consideration $\spec$ is $\monitorable{\trace}$. In this way, they can detect whether the current observed trace is ugly and inform the user. The main difference with our approach is at which level such check is performed. In their work this is performed on the property; while in our work we perform it directly on the monitor.

In~\cite{DBLP:journals/pacmpl/AcetoAFIL19,DBLP:journals/fmsd/FrancalanzaAI17}, the notion of monitorability is presented for linear and branching flavours of Hennessy-Milner Logic with recursion (recHML)~\cite{DBLP:journals/tcs/Larsen90}. Differently from us, they consider partial monitoring the monitors which derive from either safety or co-safety properties (not both). Instead, we consider the monitors which are not always capable of determining the property, either positively or negatively.

It is important to note that a property can start monitorable (resp. non-monitorable) and become non-monitorable (resp. monitorable) depending on the information known about the SUA. In fact, as pointed out in~\cite{DBLP:conf/rv/HenzingerS20}, the monitorability result w.r.t. a property may change under assumptions on the SUA. If we know how the system behaves (\textit{e.g.}, a model of the system exists), then we can rewrite the property accordingly and this can change the answer to the monitorability question. 

To the best of our knowledge, our work is the first that tackles the practical implications of partial monitoring; where monitors are not assumed to be able to always conclude the satisfaction or violation of the formal property under analysis, and where it is not always simple to determine if a property is monitorable or not.

\section{Partial Monitoring}
\label{sec:partial-monitoring}

In this section, we introduce the notion of partial monitoring from a practical perspective. First, we formally define partial monitors and show an example of how it works based on a property from a previous example. Then, we discuss the application and benefits of partial monitoring to an example of an autonomous rover performing remote inspection tasks. Finally, we present the implementation details of our tool, which is capable of automatically synthesising partial monitors from existing monitors for non-monitorable properties, and discuss the engineering aspects of our approach.

In Definition~\ref{rv-def}, we had a standard three-valued monitor. Usually, a monitor is intended to be \emph{complete}, in the sense that a verdict is always assumed to be returned. This happens due to the presence of the inconclusive verdict ($\unknown$), which is returned until the satisfaction ($\top$) or violation ($\bot$) of the property can be concluded. Nonetheless, in the standard definition, the property is assumed to be $\univmonitorable$. Moreover, most of the time these are safety properties, since RV is usually applied in scenarios where it is used to verify that ``nothing bad will ever happen''.

\begin{definition}[Partial Monitor]\label{rv-def-partial}
Let $\system$ be a system with alphabet $\eventSet$,
and $\spec$ be an LTL property. Then, a \emph{partial monitor} for $\spec$ is a function
$\stmonitor{\spec}:\eventSet^*\rightarrow\mathbb{B}_4$, where
$\mathbb{B}_4=\{\top, \bot, \unknown, \giveup \}$:
$$
\stmonitorAppl{\spec}{\trace} =
\left\{
\bgroup
\def\arraystretch{1.2}
  \begin{tabular}{cl}
  $\top$ & {\qquad$\forall_{\continuation \in \eventSet^\omega}.\trace \bullet \continuation \in \sem{\spec}$}\\
  $\bot$ & {\qquad$\forall_{\continuation \in \eventSet^\omega}.\trace \bullet \continuation \notin \sem{\spec}$}\\
  $\unknown$ & {\qquad $\exists_{\continuation\in\eventSet^*}.((\forall_{\continuation'\in\eventSet^\omega}.\trace\bullet\continuation\bullet\continuation'\in\sem{\spec})\lor(\forall_{\continuation'\in\eventSet^\omega}.\trace\bullet\continuation\bullet\continuation'\notin\sem{\spec}))$}\\
  $\giveup$ & {\qquad$otherwise$}\\
  \end{tabular}
\egroup
\right.
$$
where $\bullet$ is the standard trace concatenation operator.
\end{definition}

Definition~\ref{rv-def-partial} presents the notion of a \emph{partial monitor}, which differs from Definition~\ref{rv-def} in the values returned as outcome of the verification. An additional output ``give up'' value is added, \textit{i.e.}, $\giveup$. With $\giveup$, the monitor can explicitly give up on the current execution and inform the user/system that there is no point in continuing to monitor this property. To make the addition of this new output possible, we updated the condition for returning $\unknown$. The monitor now requires the existence of a future continuation of $\trace$ which will make the monitor conclude with a final verdict ($\top$ or $\bot$). If that is the case, then the monitor can conclude (for the moment) an inconclusive verdict, and eventually, it might conclude a final verdict. Otherwise, the monitor is unfortunately in a situation where $\trace$ denotes an ugly prefix, where no possible continuation will ever allow the monitor to conclude the satisfaction or violation of $\spec$. When this happens the monitor returns $\giveup$, which symbolises that it has given up on the current analysis.

\begin{figure}[ht]
\centering
\scalebox{0.8}{

\tikzset{every picture/.style={line width=0.75pt}} 

\begin{tikzpicture}[x=0.75pt,y=0.75pt,yscale=-1,xscale=1]

\draw  [fill={rgb, 255:red, 248; green, 231; blue, 28 }  ,fill opacity=0.7 ] (306,63) .. controls (306,49.19) and (317.19,38) .. (331,38) .. controls (344.81,38) and (356,49.19) .. (356,63) .. controls (356,76.81) and (344.81,88) .. (331,88) .. controls (317.19,88) and (306,76.81) .. (306,63) -- cycle ;
\draw  [fill={rgb, 255:red, 10; green, 100; blue, 100 }  ,fill opacity=0.3 ] (404,150) .. controls (404,136.19) and (415.19,125) .. (429,125) .. controls (442.81,125) and (454,136.19) .. (454,150) .. controls (454,163.81) and (442.81,175) .. (429,175) .. controls (415.19,175) and (404,163.81) .. (404,150) -- cycle ;
\draw  [fill={rgb, 255:red, 248; green, 231; blue, 28 }  ,fill opacity=0.7 ] (187,146) .. controls (187,132.19) and (198.19,121) .. (212,121) .. controls (225.81,121) and (237,132.19) .. (237,146) .. controls (237,159.81) and (225.81,171) .. (212,171) .. controls (198.19,171) and (187,159.81) .. (187,146) -- cycle ;
\draw  [fill={rgb, 255:red, 208; green, 2; blue, 27 }  ,fill opacity=0.7 ] (294,152) .. controls (294,138.19) and (305.19,127) .. (319,127) .. controls (332.81,127) and (344,138.19) .. (344,152) .. controls (344,165.81) and (332.81,177) .. (319,177) .. controls (305.19,177) and (294,165.81) .. (294,152) -- cycle ;
\draw  [fill={rgb, 255:red, 65; green, 117; blue, 5 }  ,fill opacity=0.7 ] (187,243) .. controls (187,229.19) and (198.19,218) .. (212,218) .. controls (225.81,218) and (237,229.19) .. (237,243) .. controls (237,256.81) and (225.81,268) .. (212,268) .. controls (198.19,268) and (187,256.81) .. (187,243) -- cycle ;
\draw    (356,63) .. controls (435.61,62.97) and (419.91,75.45) .. (428.45,122.1) ;
\draw [shift={(429,125)}, rotate = 258.74] [fill={rgb, 255:red, 0; green, 0; blue, 0 }  ][line width=0.08]  [draw opacity=0] (8.93,-4.29) -- (0,0) -- (8.93,4.29) -- cycle    ;
\draw    (429,175) .. controls (418.39,212.4) and (503.62,176.09) .. (456.26,151.13) ;
\draw [shift={(454,150)}, rotate = 385.55] [fill={rgb, 255:red, 0; green, 0; blue, 0 }  ][line width=0.08]  [draw opacity=0] (8.93,-4.29) -- (0,0) -- (8.93,4.29) -- cycle    ;
\draw    (331,88) .. controls (348.78,115.27) and (313.1,92.18) .. (318.5,124.41) ;
\draw [shift={(319,127)}, rotate = 257.84000000000003] [fill={rgb, 255:red, 0; green, 0; blue, 0 }  ][line width=0.08]  [draw opacity=0] (8.93,-4.29) -- (0,0) -- (8.93,4.29) -- cycle    ;
\draw    (319,177) .. controls (308.39,214.4) and (393.62,178.09) .. (346.26,153.13) ;
\draw [shift={(344,152)}, rotate = 385.55] [fill={rgb, 255:red, 0; green, 0; blue, 0 }  ][line width=0.08]  [draw opacity=0] (8.93,-4.29) -- (0,0) -- (8.93,4.29) -- cycle    ;
\draw    (306,63) .. controls (211.67,59.03) and (224.81,69.64) .. (212.58,118.73) ;
\draw [shift={(212,121)}, rotate = 284.54] [fill={rgb, 255:red, 0; green, 0; blue, 0 }  ][line width=0.08]  [draw opacity=0] (8.93,-4.29) -- (0,0) -- (8.93,4.29) -- cycle    ;
\draw    (212,171) .. controls (229.78,198.27) and (205.51,182.79) .. (211.48,215.4) ;
\draw [shift={(212,218)}, rotate = 257.84000000000003] [fill={rgb, 255:red, 0; green, 0; blue, 0 }  ][line width=0.08]  [draw opacity=0] (8.93,-4.29) -- (0,0) -- (8.93,4.29) -- cycle    ;
\draw    (331.23,17.97) -- (331.03,35) ;
\draw [shift={(331,38)}, rotate = 270.67] [fill={rgb, 255:red, 0; green, 0; blue, 0 }  ][line width=0.08]  [draw opacity=0] (8.93,-4.29) -- (0,0) -- (8.93,4.29) -- cycle    ;
\draw    (224,167) .. controls (213.39,204.4) and (286.97,171.97) .. (239.26,147.13) ;
\draw [shift={(237,146)}, rotate = 385.55] [fill={rgb, 255:red, 0; green, 0; blue, 0 }  ][line width=0.08]  [draw opacity=0] (8.93,-4.29) -- (0,0) -- (8.93,4.29) -- cycle    ;
\draw    (212,268) .. controls (201.39,305.4) and (286.62,269.09) .. (239.26,244.13) ;
\draw [shift={(237,243)}, rotate = 385.55] [fill={rgb, 255:red, 0; green, 0; blue, 0 }  ][line width=0.08]  [draw opacity=0] (8.93,-4.29) -- (0,0) -- (8.93,4.29) -- cycle    ;

\draw (330,63) node  [font=\large] [align=left] {?};
\draw (429,150) node  [font=\large] [align=left] {$\displaystyle \giveup$};
\draw (212,146) node  [font=\large] [align=left] {?};
\draw (319,152) node  [font=\large] [align=left] {$\displaystyle \bot $};
\draw (212,243) node  [font=\large] [align=left] {$\displaystyle \top $};
\draw (269.27,50.31) node   [align=left] {$\displaystyle ev_{1}$};
\draw (197.27,190.31) node   [align=left] {$\displaystyle ev_{2}$};
\draw (262.27,190.31) node   [align=left] {$\displaystyle *\setminus ev_{2}$};
\draw (351.28,195.31) node   [align=left] {$\displaystyle *$};
\draw (464.28,192.31) node   [align=left] {$\displaystyle *$};
\draw (408.28,52.31) node   [align=left] {$\displaystyle ev_{3}$};
\draw (295.27,94.31) node   [align=left] {$\displaystyle ev_{2} ,ev_{4}$};
\draw (250.27,284.31) node   [align=left] {$\displaystyle *$};

\end{tikzpicture}

}

\caption{Partial monitor of the $\existmonitorable$ property $\spec$ presented in Example~\ref{ex:existentiallyPZ}. Here, we can note how the previously inconclusive state on the right has now become a ``give up'' state (grey colour).}
\label{fig:partialmonitor}
\end{figure}
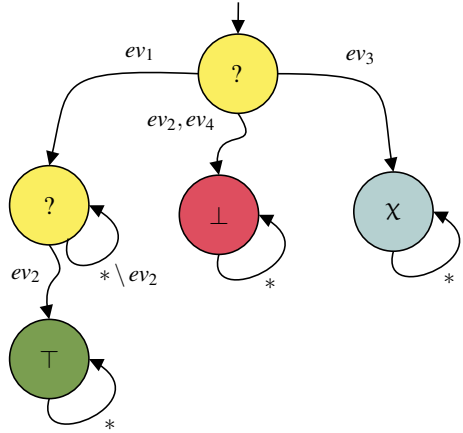

\begin{example}\label{ex:partialmonitor}
Considering once again the property of Example~\ref{ex:existentiallyPZ}, $\spec=(ev_1 \land \lozenge ev_2)\lor(ev_3 \land \square\lozenge ev_4)$, with $\eventSet=\{ev_1,ev_2,ev_3,ev_4\}$. This property is $\existmonitorable$, since not all $\trace\in\eventSet^*$ are $\monitorable{\trace}$. For this reason this property would usually be discarded, since no guarantees can be given that the resulting monitor will be able to conclude anything. In this case, by following Definition~\ref{rv-def-partial}, we can update the monitor with the additional outcome to represent the cases where it should give up.
Figure~\ref{fig:partialmonitor} reports the partial monitor obtained by updating the monitor from Figure~\ref{fig:existentiallyPZ}.
\end{example}

Given a standard monitor, to obtain a partial monitor we add an additional post-processing step after generating the Moore machine. From a Moore machine representing the instantiation of a monitor, we compute for each state labelled with $\unknown$ the reachability of a state labelled with $\top$ or $\bot$. Reaching these states means that the monitor cannot get stuck in an inconclusive state indefinitely (\textit{i.e.}, it has no dead ends). This analysis can be achieved in polynomial time w.r.t. the number of states and edges of the Moore machine. Specifically, in the worst case scenario the time complexity is $O(N^2 + N E)$, with $N$ the number of states, and $E$ the number of edges. But, to have a better understanding of the time complexity, it is important to note the relation between $N$ and $\spec$. In fact, the standard LTL monitor synthesis procedure (Figure~\ref{fsm-steps-fig}) generates a Moore machine with double exponential size w.r.t. the size of the LTL property $|\spec|$. Thus, if we expand the time complexity w.r.t. to the property, we obtain that the approach has time complexity $O(2^{2^{|\spec|+1}} + 2^{2^{|\spec|}} E)$. Nonetheless, properties are usually small in size, which makes the time complexity less limiting.

\subsection{A Remote Inspection Example}
\label{sec:example}

We demonstrate the usefulness of our approach by applying it to the remote inspection example mentioned in the introduction. This example is based on a simulation, first introduced in~\cite{robotics10030086}, of an autonomous rover deployed to perform remote inspection of nuclear facilities. The rover has access to sensors, which are used to detect the level of radiation, and a camera, which is used to acquire images of tanks (containing radioactive material) to perform an integrity analysis (\textit{e.g.}, deterioration of the container). The objective of the rover is to patrol and inspect important locations (\textit{i.e.}, waypoints) around the facility. As part of the inspection of a particular location, the rover has to take measurements of the radiation level when it arrives in such location.


\begin{example}\label{ex:casestudy1}
We start by demonstrating our approach applied to this example with a very simple property $\spec = \square\lozenge inspect\_tank\_1$. This property is shown in Figure~\ref{fig:simple}, with Figure~\ref{fig:simple1} containing the traditional Moore machine, and then after applying our technique we can see in Figure~\ref{fig:simple2} that the monitor can only give up in this case. Intuitively, this property states that it is always the case that eventually the rover will inspect the waypoint \emph{tank1}. Since the rover has to constantly patrol these waypoints, it makes sense to represent this behaviour with such property. However, we note that there are many other ways to write this property, and some may sacrifice generalisation to write a property that is monitorable. That is a valid approach, and for simple cases such as with this property it is indeed the best solution, since after applying our approach we ended up with a monitor that is only able to give up (\textit{i.e.}, no partial monitoring is possible).
\end{example}

\begin{figure}[ht]
	\centering
	\begin{subfigure}[t]{0.46\textwidth}
        \centering
\centering
\scalebox{0.8}{

\tikzset{every picture/.style={line width=0.75pt}} 

\begin{tikzpicture}[x=0.75pt,y=0.75pt,yscale=-1,xscale=1]

\draw  [fill={rgb, 255:red, 248; green, 231; blue, 28 }  ,fill opacity=0.7 ] (199,73) .. controls (199,59.19) and (210.19,48) .. (224,48) .. controls (237.81,48) and (249,59.19) .. (249,73) .. controls (249,86.81) and (237.81,98) .. (224,98) .. controls (210.19,98) and (199,86.81) .. (199,73) -- cycle ;
\draw    (224.23,27.97) -- (224.03,45) ;
\draw [shift={(224,48)}, rotate = 270.67] [fill={rgb, 255:red, 0; green, 0; blue, 0 }  ][line width=0.08]  [draw opacity=0] (8.93,-4.29) -- (0,0) -- (8.93,4.29) -- cycle    ;
\draw    (236,94) .. controls (225.39,131.4) and (298.97,98.97) .. (251.26,74.13) ;
\draw [shift={(249,73)}, rotate = 385.55] [fill={rgb, 255:red, 0; green, 0; blue, 0 }  ][line width=0.08]  [draw opacity=0] (8.93,-4.29) -- (0,0) -- (8.93,4.29) -- cycle    ;

\draw (224,73) node  [font=\large] [align=left] {?};
\draw (268.27,111.31) node   [align=left] {$\displaystyle *$};

\end{tikzpicture}

}

\caption{Traditional Moore machine monitor generated from property $\spec$.}
\label{fig:simple1}
    \end{subfigure}%
    \hspace{0.3cm}
    \begin{subfigure}[t]{0.46\textwidth}
        \centering
\centering
\scalebox{0.8}{

\tikzset{every picture/.style={line width=0.75pt}} 

\begin{tikzpicture}[x=0.75pt,y=0.75pt,yscale=-1,xscale=1]

\draw  [fill={rgb, 255:red, 10; green, 100; blue, 100 }  ,fill opacity=0.3 ] (199,73) .. controls (199,59.19) and (210.19,48) .. (224,48) .. controls (237.81,48) and (249,59.19) .. (249,73) .. controls (249,86.81) and (237.81,98) .. (224,98) .. controls (210.19,98) and (199,86.81) .. (199,73) -- cycle ;
\draw    (224.23,27.97) -- (224.03,45) ;
\draw [shift={(224,48)}, rotate = 270.67] [fill={rgb, 255:red, 0; green, 0; blue, 0 }  ][line width=0.08]  [draw opacity=0] (8.93,-4.29) -- (0,0) -- (8.93,4.29) -- cycle    ;
\draw    (236,94) .. controls (225.39,131.4) and (298.97,98.97) .. (251.26,74.13) ;
\draw [shift={(249,73)}, rotate = 385.55] [fill={rgb, 255:red, 0; green, 0; blue, 0 }  ][line width=0.08]  [draw opacity=0] (8.93,-4.29) -- (0,0) -- (8.93,4.29) -- cycle    ;

\draw (224,73) node  [font=\large] [align=left] {$\giveup$};
\draw (268.27,111.31) node   [align=left] {$\displaystyle *$};

\end{tikzpicture}

}

\caption{New monitor generated after applying our approach.}
\label{fig:simple2}
    \end{subfigure}
     \caption{\label{fig:simple}A simple property $\spec = \square\lozenge inspect\_tank\_1$.}
\end{figure}

\begin{example}\label{ex:casestudy2}
Next, let us consider a more interesting property where we can demonstrate that partial monitoring can indeed be useful:
\begin{eqnarray*}
\spec = radiation\_low \; \until ((radiation\_high \land \lozenge move\_to\_decontamination) \lor \\
(radiation\_medium \land \square\lozenge (inspect\_tank\_1 \lor inspect\_tank\_2 \lor \ldots \lor inspect\_tank\_n)))
\end{eqnarray*}
This property says that we can observe the event radiation\_low until we observe either radiation\_high or radiation\_medium. Low, medium, and high radiation refer to the level of radiation that is currently observed by the radiation sensor. If radiation\_high is observed, then eventually we have to observe the event move\_to\_decontamination, which represents the command being sent to move the rover to a decontamination zone (\textit{i.e.}, high level of radiation can be dangerous to the rover). Otherwise, if we observe radiation\_medium, then we have to patrol one of the radiation tanks ($1 \ldots n$) to identify if there are any abnormalities.
\end{example}

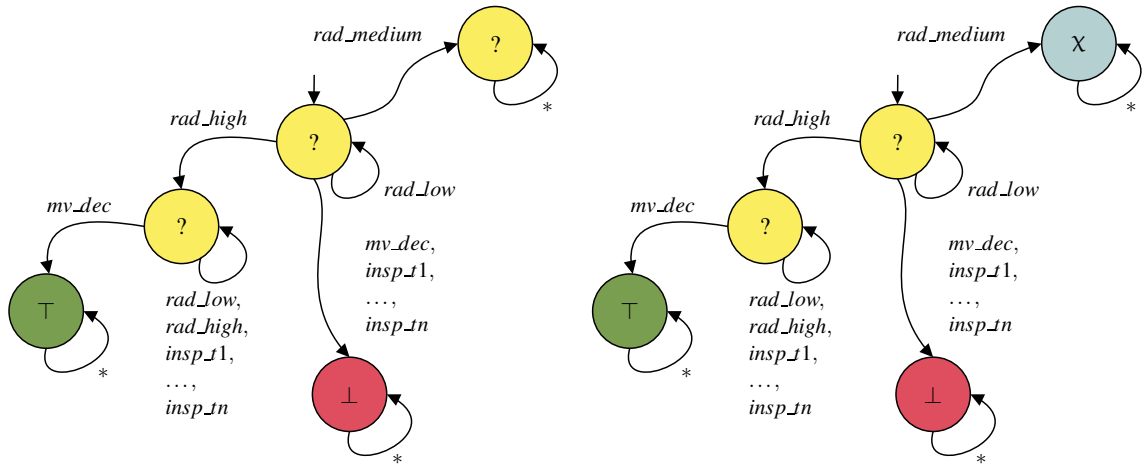
\begin{figure}[ht]
	\centering
	\begin{subfigure}[t]{0.46\textwidth}
        \centering
\centering
\scalebox{0.75}{

\tikzset{every picture/.style={line width=0.75pt}} 

\begin{tikzpicture}[x=0.75pt,y=0.75pt,yscale=-1,xscale=1]

\draw  [fill={rgb, 255:red, 248; green, 231; blue, 28 }  ,fill opacity=0.7 ] (385,66) .. controls (385,52.19) and (396.19,41) .. (410,41) .. controls (423.81,41) and (435,52.19) .. (435,66) .. controls (435,79.81) and (423.81,91) .. (410,91) .. controls (396.19,91) and (385,79.81) .. (385,66) -- cycle ;
\draw  [fill={rgb, 255:red, 248; green, 231; blue, 28 }  ,fill opacity=0.7 ] (174,189) .. controls (174,175.19) and (185.19,164) .. (199,164) .. controls (212.81,164) and (224,175.19) .. (224,189) .. controls (224,202.81) and (212.81,214) .. (199,214) .. controls (185.19,214) and (174,202.81) .. (174,189) -- cycle ;
\draw  [fill={rgb, 255:red, 208; green, 2; blue, 27 }  ,fill opacity=0.7 ] (287,302) .. controls (287,288.19) and (298.19,277) .. (312,277) .. controls (325.81,277) and (337,288.19) .. (337,302) .. controls (337,315.81) and (325.81,327) .. (312,327) .. controls (298.19,327) and (287,315.81) .. (287,302) -- cycle ;
\draw  [fill={rgb, 255:red, 65; green, 117; blue, 5 }  ,fill opacity=0.7 ] (83,246) .. controls (83,232.19) and (94.19,221) .. (108,221) .. controls (121.81,221) and (133,232.19) .. (133,246) .. controls (133,259.81) and (121.81,271) .. (108,271) .. controls (94.19,271) and (83,259.81) .. (83,246) -- cycle ;
\draw    (410,91) .. controls (399.39,128.4) and (484.62,92.09) .. (437.26,67.13) ;
\draw [shift={(435,66)}, rotate = 385.55] [fill={rgb, 255:red, 0; green, 0; blue, 0 }  ][line width=0.08]  [draw opacity=0] (8.93,-4.29) -- (0,0) -- (8.93,4.29) -- cycle    ;
\draw    (312,327) .. controls (301.39,364.4) and (386.62,328.09) .. (339.26,303.13) ;
\draw [shift={(337,302)}, rotate = 385.55] [fill={rgb, 255:red, 0; green, 0; blue, 0 }  ][line width=0.08]  [draw opacity=0] (8.93,-4.29) -- (0,0) -- (8.93,4.29) -- cycle    ;
\draw    (288.23,86.97) -- (288.03,104) ;
\draw [shift={(288,107)}, rotate = 270.67] [fill={rgb, 255:red, 0; green, 0; blue, 0 }  ][line width=0.08]  [draw opacity=0] (8.93,-4.29) -- (0,0) -- (8.93,4.29) -- cycle    ;
\draw    (108,271) .. controls (97.39,308.4) and (182.62,272.09) .. (135.26,247.13) ;
\draw [shift={(133,246)}, rotate = 385.55] [fill={rgb, 255:red, 0; green, 0; blue, 0 }  ][line width=0.08]  [draw opacity=0] (8.93,-4.29) -- (0,0) -- (8.93,4.29) -- cycle    ;
\draw  [fill={rgb, 255:red, 248; green, 231; blue, 28 }  ,fill opacity=0.7 ] (263,132) .. controls (263,118.19) and (274.19,107) .. (288,107) .. controls (301.81,107) and (313,118.19) .. (313,132) .. controls (313,145.81) and (301.81,157) .. (288,157) .. controls (274.19,157) and (263,145.81) .. (263,132) -- cycle ;
\draw    (301,153) .. controls (290.39,190.4) and (363,157.97) .. (315.27,133.13) ;
\draw [shift={(313,132)}, rotate = 385.55] [fill={rgb, 255:red, 0; green, 0; blue, 0 }  ][line width=0.08]  [draw opacity=0] (8.93,-4.29) -- (0,0) -- (8.93,4.29) -- cycle    ;
\draw    (263,132) .. controls (222.05,125.18) and (192.73,127.82) .. (198.49,161.36) ;
\draw [shift={(199,164)}, rotate = 257.84000000000003] [fill={rgb, 255:red, 0; green, 0; blue, 0 }  ][line width=0.08]  [draw opacity=0] (8.93,-4.29) -- (0,0) -- (8.93,4.29) -- cycle    ;
\draw    (288,157) .. controls (305.96,184.55) and (271.07,231.56) .. (310.16,275.02) ;
\draw [shift={(312,277)}, rotate = 226.32999999999998] [fill={rgb, 255:red, 0; green, 0; blue, 0 }  ][line width=0.08]  [draw opacity=0] (8.93,-4.29) -- (0,0) -- (8.93,4.29) -- cycle    ;
\draw    (308,117) .. controls (368.09,106.17) and (321.45,86.6) .. (382.15,66.9) ;
\draw [shift={(385,66)}, rotate = 522.9] [fill={rgb, 255:red, 0; green, 0; blue, 0 }  ][line width=0.08]  [draw opacity=0] (8.93,-4.29) -- (0,0) -- (8.93,4.29) -- cycle    ;
\draw    (212,210) .. controls (201.39,247.4) and (274,214.97) .. (226.27,190.13) ;
\draw [shift={(224,189)}, rotate = 385.55] [fill={rgb, 255:red, 0; green, 0; blue, 0 }  ][line width=0.08]  [draw opacity=0] (8.93,-4.29) -- (0,0) -- (8.93,4.29) -- cycle    ;
\draw    (174,189) .. controls (133.05,182.18) and (103.73,184.82) .. (109.49,218.36) ;
\draw [shift={(110,221)}, rotate = 257.84000000000003] [fill={rgb, 255:red, 0; green, 0; blue, 0 }  ][line width=0.08]  [draw opacity=0] (8.93,-4.29) -- (0,0) -- (8.93,4.29) -- cycle    ;

\draw (410,66) node  [font=\large] [align=left] {?};
\draw (199,189) node  [font=\large] [align=left] {?};
\draw (312,302) node  [font=\large] [align=left] {$\displaystyle \bot $};
\draw (108,246) node  [font=\large] [align=left] {$\displaystyle \top $};
\draw (344.28,345.31) node   [align=left] {$\displaystyle *$};
\draw (445.28,108.31) node   [align=left] {$\displaystyle *$};
\draw (146.27,287.31) node   [align=left] {$\displaystyle *$};
\draw (288,132) node  [font=\large] [align=left] {?};
\draw (359.28,163.31) node   [align=left] {$\displaystyle rad\_low$};
\draw (217.27,117.31) node   [align=left] {$\displaystyle rad\_high$};
\draw (346.28,228.31) node   [align=left] {$\displaystyle  \begin{array}{{>{\displaystyle}l}}
mv\_dec,\ \\
insp\_t1,\\
\ldots ,\\
insp\_tn
\end{array}$};
\draw (324.27,59.31) node   [align=left] {$\displaystyle rad\_medium$};
\draw (216.27,275.31) node   [align=left] {$\displaystyle  \begin{array}{{>{\displaystyle}l}}
rad\_low,\\
rad\_high,\\
insp\_t1,\\
\ldots ,\\
insp\_tn
\end{array}$};
\draw (130.27,174.31) node   [align=left] {$\displaystyle mv\_dec$};

\end{tikzpicture}

}

\caption{Monitor (as Moore machine) for the property.}
\label{fig:case-study-fsm1}
    \end{subfigure}%
    \hspace{0.3cm}
    \begin{subfigure}[t]{0.46\textwidth}
        \centering
\centering
\scalebox{0.75}{

\tikzset{every picture/.style={line width=0.75pt}} 

\begin{tikzpicture}[x=0.75pt,y=0.75pt,yscale=-1,xscale=1]

\draw  [fill={rgb, 255:red, 10; green, 100; blue, 100 }  ,fill opacity=0.3 ] (385,66) .. controls (385,52.19) and (396.19,41) .. (410,41) .. controls (423.81,41) and (435,52.19) .. (435,66) .. controls (435,79.81) and (423.81,91) .. (410,91) .. controls (396.19,91) and (385,79.81) .. (385,66) -- cycle ;
\draw  [fill={rgb, 255:red, 248; green, 231; blue, 28 }  ,fill opacity=0.7 ] (174,189) .. controls (174,175.19) and (185.19,164) .. (199,164) .. controls (212.81,164) and (224,175.19) .. (224,189) .. controls (224,202.81) and (212.81,214) .. (199,214) .. controls (185.19,214) and (174,202.81) .. (174,189) -- cycle ;
\draw  [fill={rgb, 255:red, 208; green, 2; blue, 27 }  ,fill opacity=0.7 ] (287,302) .. controls (287,288.19) and (298.19,277) .. (312,277) .. controls (325.81,277) and (337,288.19) .. (337,302) .. controls (337,315.81) and (325.81,327) .. (312,327) .. controls (298.19,327) and (287,315.81) .. (287,302) -- cycle ;
\draw  [fill={rgb, 255:red, 65; green, 117; blue, 5 }  ,fill opacity=0.7 ] (83,246) .. controls (83,232.19) and (94.19,221) .. (108,221) .. controls (121.81,221) and (133,232.19) .. (133,246) .. controls (133,259.81) and (121.81,271) .. (108,271) .. controls (94.19,271) and (83,259.81) .. (83,246) -- cycle ;
\draw    (410,91) .. controls (399.39,128.4) and (484.62,92.09) .. (437.26,67.13) ;
\draw [shift={(435,66)}, rotate = 385.55] [fill={rgb, 255:red, 0; green, 0; blue, 0 }  ][line width=0.08]  [draw opacity=0] (8.93,-4.29) -- (0,0) -- (8.93,4.29) -- cycle    ;
\draw    (312,327) .. controls (301.39,364.4) and (386.62,328.09) .. (339.26,303.13) ;
\draw [shift={(337,302)}, rotate = 385.55] [fill={rgb, 255:red, 0; green, 0; blue, 0 }  ][line width=0.08]  [draw opacity=0] (8.93,-4.29) -- (0,0) -- (8.93,4.29) -- cycle    ;
\draw    (288.23,86.97) -- (288.03,104) ;
\draw [shift={(288,107)}, rotate = 270.67] [fill={rgb, 255:red, 0; green, 0; blue, 0 }  ][line width=0.08]  [draw opacity=0] (8.93,-4.29) -- (0,0) -- (8.93,4.29) -- cycle    ;
\draw    (108,271) .. controls (97.39,308.4) and (182.62,272.09) .. (135.26,247.13) ;
\draw [shift={(133,246)}, rotate = 385.55] [fill={rgb, 255:red, 0; green, 0; blue, 0 }  ][line width=0.08]  [draw opacity=0] (8.93,-4.29) -- (0,0) -- (8.93,4.29) -- cycle    ;
\draw  [fill={rgb, 255:red, 248; green, 231; blue, 28 }  ,fill opacity=0.7 ] (263,132) .. controls (263,118.19) and (274.19,107) .. (288,107) .. controls (301.81,107) and (313,118.19) .. (313,132) .. controls (313,145.81) and (301.81,157) .. (288,157) .. controls (274.19,157) and (263,145.81) .. (263,132) -- cycle ;
\draw    (301,153) .. controls (290.39,190.4) and (363,157.97) .. (315.27,133.13) ;
\draw [shift={(313,132)}, rotate = 385.55] [fill={rgb, 255:red, 0; green, 0; blue, 0 }  ][line width=0.08]  [draw opacity=0] (8.93,-4.29) -- (0,0) -- (8.93,4.29) -- cycle    ;
\draw    (263,132) .. controls (222.05,125.18) and (192.73,127.82) .. (198.49,161.36) ;
\draw [shift={(199,164)}, rotate = 257.84000000000003] [fill={rgb, 255:red, 0; green, 0; blue, 0 }  ][line width=0.08]  [draw opacity=0] (8.93,-4.29) -- (0,0) -- (8.93,4.29) -- cycle    ;
\draw    (288,157) .. controls (305.96,184.55) and (271.07,231.56) .. (310.16,275.02) ;
\draw [shift={(312,277)}, rotate = 226.32999999999998] [fill={rgb, 255:red, 0; green, 0; blue, 0 }  ][line width=0.08]  [draw opacity=0] (8.93,-4.29) -- (0,0) -- (8.93,4.29) -- cycle    ;
\draw    (308,117) .. controls (368.09,106.17) and (321.45,86.6) .. (382.15,66.9) ;
\draw [shift={(385,66)}, rotate = 522.9] [fill={rgb, 255:red, 0; green, 0; blue, 0 }  ][line width=0.08]  [draw opacity=0] (8.93,-4.29) -- (0,0) -- (8.93,4.29) -- cycle    ;
\draw    (212,210) .. controls (201.39,247.4) and (274,214.97) .. (226.27,190.13) ;
\draw [shift={(224,189)}, rotate = 385.55] [fill={rgb, 255:red, 0; green, 0; blue, 0 }  ][line width=0.08]  [draw opacity=0] (8.93,-4.29) -- (0,0) -- (8.93,4.29) -- cycle    ;
\draw    (174,189) .. controls (133.05,182.18) and (103.73,184.82) .. (109.49,218.36) ;
\draw [shift={(110,221)}, rotate = 257.84000000000003] [fill={rgb, 255:red, 0; green, 0; blue, 0 }  ][line width=0.08]  [draw opacity=0] (8.93,-4.29) -- (0,0) -- (8.93,4.29) -- cycle    ;

\draw (410,66) node  [font=\large] [align=left] {$\giveup$};
\draw (199,189) node  [font=\large] [align=left] {?};
\draw (312,302) node  [font=\large] [align=left] {$\displaystyle \bot $};
\draw (108,246) node  [font=\large] [align=left] {$\displaystyle \top $};
\draw (344.28,345.31) node   [align=left] {$\displaystyle *$};
\draw (445.28,108.31) node   [align=left] {$\displaystyle *$};
\draw (146.27,287.31) node   [align=left] {$\displaystyle *$};
\draw (288,132) node  [font=\large] [align=left] {?};
\draw (359.28,163.31) node   [align=left] {$\displaystyle rad\_low$};
\draw (217.27,117.31) node   [align=left] {$\displaystyle rad\_high$};
\draw (346.28,228.31) node   [align=left] {$\displaystyle  \begin{array}{{>{\displaystyle}l}}
mv\_dec,\ \\
insp\_t1,\\
\ldots ,\\
insp\_tn
\end{array}$};
\draw (324.27,59.31) node   [align=left] {$\displaystyle rad\_medium$};
\draw (216.27,275.31) node   [align=left] {$\displaystyle  \begin{array}{{>{\displaystyle}l}}
rad\_low,\\
rad\_high,\\
insp\_t1,\\
\ldots ,\\
insp\_tn
\end{array}$};
\draw (130.27,174.31) node   [align=left] {$\displaystyle mv\_dec$};

\end{tikzpicture}

}

\caption{Updated monitor after applying our technique.}
\label{fig:case-study-fsm2}
    \end{subfigure}
     \caption{\label{fig:notsimple}A property that deals with the different levels of radiation. \emph{rad} is short for radiation, \emph{insp} is short for inspection, \emph{t} is short for tank, and \emph{mv\_dec} is short for move to decontamination.}
\end{figure}

The monitor (represented as a Moore machine) for this more complex property is shown in Figure~\ref{fig:notsimple}. This monitor originally has three inconclusive states, as shown in Figure~\ref{fig:case-study-fsm1}. The first is the initial state which will stay inconclusive when it observes \emph{rad\_low}, until it observes \emph{rad\_high} and then moves to the left branch, or \emph{rad\_medium} and then moves to the right branch, or any other event and then moves to the centre branch. Since the initial state is inconclusive, we have to expand it to look for positive and negative states. The centre branch is immediately expanded into a negative state, thus, we know that the initial state should remain inconclusive (\textit{i.e.}, not output give up). Looking at the left branch, we encounter the second inconclusive state, but we can quickly notice that upon observing \emph{mv\_dec} we arrive in a positive state, thus, we also know that the inconclusive state in the left branch should remain inconclusive. Finally, the third and last inconclusive state can be found in the right branch. There is no transition from this state to any other state that leads into positive or negative states, therefore, this state should output give up. Figure~\ref{fig:case-study-fsm2} contains the result of this reachability analysis.

\subsection{Implementation}
\label{sec:implementation}

We made a Java program\footnote{\url{https://github.com/AngeloFerrando/PartialMonitor}} which implements the transformation from standard to partial monitor. This tool depends on LamaConv\footnote{\url{https://www.isp.uni-luebeck.de/lamaconv}}, a Java library that is capable of translating expressions in temporal logics into equivalent automata, and to generate monitors out of these automata. These monitors can then be used for runtime verification and/or model checking. Our tool calls LamaConv and makes it generate a standard three-valued LTL monitor. After that, for each inconclusive state it performs a reachability analysis. Each inconclusive state that cannot reach any non-inconclusive state (\textit{i.e.}, $\top$ or $\bot$ labelled states) is then labelled with $\giveup$ instead of $\unknown$. In this way, monitors can still be generated and used for partial monitoring of non-monitorable properties, since ugly prefixes are explicitly recognised and the monitoring process consequently interrupted.

From an engineering perspective, our tool takes as input an LTL property and then provides a human-readable output. To be more precise, three input parameters have to be set when executing our tool:
\begin{enumerate}
    \item the path to the folder containing LamaConv installation (where ``\emph{rltlconv.jar}'' can be found);
    \item $\spec$, the LTL property that will be used for synthesising the monitor (standard LTL syntax as accepted by LamaConv);
    \item $\eventSet$, the alphabet of the SUA. 
\end{enumerate}

Given the input described above, our tool starts by calling LamaConv, which generates a character string in the intuitive Automata File Format (AFF) format. This is then parsed by our tool into a Java object, and the reachability analysis is performed to detect $\giveup$ (give up) states. 

Our tool can generate two outputs, the updated AFF and/or a Java object. The AFF can be directly read and processed as a string representation of the monitor. When parsing this updated AFF into an application, the user should be aware that the only change that our tool makes to the AFF is to update the inconclusive states that were identified to never conclude the satisfaction or violation of the property with a give up symbol (by replacing ``?'' with ``x'' in the AFF). Otherwise, if using the Java object, then the tool can be included directly as an external Java library, and the monitor can be used as a Java object by calling the available methods. In this way, the monitor generated by the tool can be easily integrated with third-party software. 

\section{Discussion}
\label{sec:discussion}

In this work, we presented an intuitive approach to make monitors capable of giving up when necessary. As mentioned in Section~\ref{sec:monitorability}, this is not the first time the notion of monitorability has been studied. Nonetheless, w.r.t. related work, we tackle the monitorability problem on a more practical level. Indeed, there are plenty of works which study and explore the theoretical aspects of what makes a property monitorable; but, not much has been done to answer what we can do with monitorability in practice. For instance, when is it the right time to give up on a property? Or more generally, what can we do with a non-monitorable property? Naturally, there are scenarios where nothing can be done. These are the cases when a property simply cannot be verified at runtime, in any possible way (such as the property from Example~\ref{ex:casestudy1}). However, there are scenarios where something can be rightfully concluded, albeit partially. And these are the cases we aimed to exploit in this work. In general, properties are expected to be fully monitorable (\textit{i.e.}, $\univmonitorable$), because when such constraint does not hold, we do not have guarantees whether the monitor will ever conclude anything useful. Nonetheless, if the monitor is capable of giving up by recognising and handling ugly prefixes, then non-monitorable properties can be monitored through the use of partial monitors.

Applying such analysis at the monitor level is very important, because this does not only allow us to give up on the monitor at runtime, but also to reuse our approach in various scenarios. Since the approach is based on the Moore machine denoting the monitor (and not the property), it is formalism-agnostic up to a certain level. Thus, we are not just limited to LTL for defining the properties that can be used. We can use another logic as long as a Moore machine can be synthesised. This would not require any modifications to our approach at the theoretical level, but it does require changes in the implementation. For example, either the logic can be converted into LTL if possible, or an automaton representing the monitor needs to be generated (if this automaton is a Moore machine we are done and ready to use our approach, otherwise we need to convert it). 

From a research perspective, by directly applying our approach on a Moore machine, we also offer a much more reusable workflow. As long as a Moore machine is generated, more challenging aspects can be explored. For instance, Predictive Runtime Verification (PRV)~\cite{DBLP:conf/nfm/ZhangLD12,DBLP:conf/rv/Leucker12} can be deployed instead of standard RV. In fact, works on PRV of LTL properties exist; where a model of the system (a B\"{u}chi Automaton) is used to predict future events and to help the monitor to conclude its verdict in advance (before actually observing the events). In such works, the flow presented in Figure~\ref{fsm-steps-fig} is extended for considering the model of the system as well. The important aspect for us is that, even though the workflow is extended, the final result is still a Moore machine, with the additional power of anticipating the conclusive outcomes. Since our approach is directly applied on a Moore machine and not on the property itself, we can still obtain partial monitoring for PRV by analysing the resulting predictive Moore machine. This means we can apply our approach to more challenging scenarios in the future, without the need to change anything specific in the process.

To summarise, in this paper we introduced the notion of partial monitoring as a practical view on monitorability, but it is important to note that the theoretical aspects of partial monitorability are different and much harder to tackle. On one hand we have partial monitoring, where we look into the representation of an existing monitor and we identify if it has any states where it should give up; if this is the case and the monitor still has other valid states that are not give up, then we have a partial monitor. Partial monitorability on the other hand, would deal with identifying what can make a property partially monitorable; \textit{e.g.,} what is the relation of the chosen logic's operators with monitorability (if any) and how chaining these operators together impacts monitorability, delineate the types of properties that are more amenable and advantageous for partial monitoring, and so on.

\section{Conclusions and Future Work}
\label{sec:conclusions}

In this paper, we discussed the issue of handling monitors generated from non-monitorable properties and how such monitors can be extended to give up when no final verdict can be ever concluded. We described a practical technique to perform reachability analysis of LTL monitors obtained using standard synthesis approaches~\cite{10.1145/2000799.2000800}, and how the resulting \emph{partial monitors} can avoid to get stuck in scenarios where no final verdict can be concluded, such as when trying to monitor non-monitorable properties. We have demonstrated the use of partial monitoring in an example from the robotics domain, where a rover performing the inspection of a nuclear facility can be partially verified at runtime using non-monitorable properties. We also described the implementation details and engineering aspects of a tool that we developed to automate the detection and synthesis of partial monitors.

This work has focused on the engineering and implications of monitorability for the synthesis of partial monitors. In future work, we plan to extend the approach to other formalisms, such as Metric Temporal Logic (MTL)~\cite{DBLP:journals/rts/Koymans90}, Signal Temporal Logic (STL)~\cite{DBLP:conf/formats/MalerN04}, Runtime Monitoring Language (RML)~\cite{DBLP:journals/scp/AnconaFFM21}, and so on. We also want to integrate our approach with existing RV tools for autonomous systems such as ROSMonitoring~\cite{Ferrando20a}, which is a RV framework for the popular middleware for robotic development Robot Operating System (ROS)\footnote{\url{https://www.ros.org/}}.
Adding support to other formalisms requires updates to the tool, since right now it can only take as input plain Moore machines. This is important since more complex scenarios may require more expressive formalisms than LTL, and the issue of having partial monitors (as well as monitorability) arises in other formalisms as well.

\bibliographystyle{eptcs}
\bibliography{main}
\end{document}